\pdfoutput=1
\documentclass[apj,twocolumn]{emulateapj}
\usepackage{natbib}
\usepackage{booktabs,graphicx,amsmath,paralist,longtable,hyperref}
\usepackage[version=4]{mhchem}
\usepackage{dcolumn,xcolor,braket,cleveref,rotating}

\newcommand{\Marvel}{{\sc Marvel}}
\newcommand{\ZrO}{\ce{^{90}Zr^{16}O}}
\newcommand{\cm}{cm$^{-1}$}

\newcommand{\X}{X $^1\Sigma^+$}
\newcommand{\A}{A $^1\Delta$}
\newcommand{\B}{B $^1\Pi$}
\newcommand{\C}{C $^1\Sigma^+$}
\newcommand{\D}{D $^1\Gamma$}
\newcommand{\E}{E $^1\Phi$}
\newcommand{\F}{F $^1\Delta$}

\usepackage{array}
\newcolumntype{H}{>{\setbox0=\hbox\bgroup}c<{\egroup}@{}}
\newcommand{\Ta}{a $^3\Delta$}
\newcommand{\Tb}{b $^3\Pi$}
\newcommand{\Tc}{c $^3\Sigma^-$}
\newcommand{\Td}{d $^3\Phi$}
\newcommand{\Te}{e $^3\Pi$}
\newcommand{\Tf}{f $^3\Delta$}

\newcommand{\mc}{\multicolumn}
\newcolumntype{d}{D{.}{.}{-1}}
\newcommand{\alert}[1]{}
\newcommand{\red}[1]{}

\newcommand{\noenergy}{8088}
\newcommand{\notrans}{23~317}
\newcommand{\novalid}{22~549}
\newcommand{\noelec}{9}
\newcommand{\nospinvibronic}{72}

\begin{document}

\title{\Marvel\ analysis of the measured high-resolution rovibronic spectra of \ZrO}

\author{Laura K McKemmish$^{1,2}$, Jasmin Borsovszky$^{2}$, Katie L Goodhew$^3$, Samuel Sheppard$^3$,  Aphra F V Bennett$^3$, Alfie D J Martin$^3$, Amrik Singh$^3$, Callum A J Sturgeon$^3$,  Tibor Furtenbacher$^4$, Attila G. Cs\'asz\'ar$^4$, Jonathan Tennyson$^1$}
\affil{$^1$Department of Physics and Astronomy, University College London, London, WC1E 6BT, UK \\$^2$ School of Chemistry, University of New South Wales, Kensington, Sydney, Australia\\
$^3$Highams Park School, Handsworth Avenue, Highams Park, London, E4 9PJ, UK \\
$^4$Institute  of Chemistry, 
Lor\'and E\"otv\"os University and MTA-ELTE Complex Chemical Systems Research Group,
H-1518 Budapest 112, Hungary}
\email{laura.mckemmish@gmail.com}

\begin{abstract}
  Zirconium oxide(ZrO) is an important astrophysical molecule that
  defines the S-star classification class for cool giant stars.  Accurate,
  empirical rovibronic energy levels, with associated labels and
  uncertainties, are reported for \noelec{} low-lying electronic states of the
  diatomic \ZrO\ molecule.
These \noenergy{} empirical energy levels are determined using the \Marvel\ 
(Measured Active 
Rotational-Vibrational Energy Levels) algorithm with \notrans{} input assigned 
transition frequencies, \novalid{} of which were validated. 
A temperature-dependent partition function is presented alongside
updated spectroscopic constants for the \noelec{} low-lying electronic states.
\end{abstract}


\section{Introduction}

\renewcommand{\arraystretch}{1.1}

ZrO is a transition metal diatomic oxide which, like similar species, possesses strong absorption lines and a complex electronic structure. Strong  ZrO absorption lines are the identifying characteristic of the rare $S$-type stars \citep{22Merrill,54Keenan.Sstars,78WyCl.Sstars,79Ak.Stars,80KeBo.Sstars,88LiLi.Sstars,00VaJo.Sstars}. Traditionally thought to be caused by carbon/oxygen ratios near unity \citep{79Ak.Stars,86SmVeLa.Sstars}, the recent investigation by \citet{17VaNeJo.ZrO} confirms the earlier claim by \cite{80Pi.ZrO} that the ZrO lines are caused by overabundance of s elements like Zr. Weak ZrO bands are characteristic of SC stars \citep{80KeBo.Sstars,04ZiBeMa.Sstars}. Faint ZrO bands have also been identified in sun-spots \citep{31Richardson.ZrO,12SrSh.ZrO} and M-stars \citep{34Bobrov.ZrO}.

The ZrO absorption bands were first observed in spectra taken by
\citet{22Merrill}, with \citet{24King.ZrO} providing laboratory
confirmation of the molecular origin of the bands.
\citet{54Keenan.Sstars} provided the first classification of S-type
stars. Early studies of ZrO bands in stars include an analysis of R
Geminorum by \citet{55Phillips.ZrO}.

The presence of ZrO (and other s-process elements) in S-stars is due
to the nucleosynthesis s-process occurring within these stars
\citep{98JoHiWa.Sstars} or in a companion star before being accreted to their surface \citep{00VaJo.Sstars}. The s-process only occurs at relatively low
neutron densities and intermediate temperature conditions.  There are
two types of S-stars depending on whether the s-process elements are
formed within the star itself or transferred from a binary partner
star. Short-lived cooler intrinsic S-stars are formed in around 10\%
of asymptotic giant branch stars when s-process elements convect to
the surface due to dredge-up during the short thermal pulse-asymptotic
giant branch phase \citep{85SmLa.Sstars,00VaJo.Sstars}. Longer-lived
hotter extrinsic stars are formed due to binary system mass transfer
\citep{95LaSmVe.Sstars,00VaJo.Sstars}, and are evolutionary understood
as the descendants of barium stars \citep{00VaJo.Sstars}. They can be
distinguished by the presence of Tc in intrinsic S-stars
\citep{99VaJo.Sstars,00VaJoUd.Sstars,00VaJo.Sstars}. 

\citet{85LiDa.ZrO} are regularly cited as providing 330,000 lines of a
ZrO line list; however, these data are not available as part of the original
publcation. It is likely this cited line list consists of model
Hamiltonian fits to the main bands along with band intensities,
Franck-Condon factors and H\"onl-London factors. This has been
superseded by the line list using similar methods created by \citet{PlezZrO}, which is unpublished but freely available
online. 
There is thus, to our knowledge, no available line list
created using variational nuclear motion methods from fitted potential
energy, {\it ab initio} dipole moment and fitted spin-orbit coupling curves,
as can be constructed using current techniques by, e.g. the ExoMol
group \citep{jt693}. Such studies are greatly aided by the availability
of accurate empirical energy levels such as the dataset developed in this paper. 


Due to its astrophysical importance, ZrO has been the subject of a
large number of experimental studies. One of the aims of this paper is to
review and compile the spectroscopic data from these previous studies
to produce a single recommended list of experimentally derived
empirical energy levels and validated transition frequencies.
As part of this process, we  extracted all
previous experimental data into a consistent set of assigned
transition frequencies with uncertainties. Future experimental results
can be added to this Master List to obtain an updated list of
empirical energy levels using the \textsc{Marvel} program(described
below). We anticipate that these energy levels will be used to refine
new spectroscopic models for \ZrO\ and produce updated extensive hot
molecular line lists for use in atmospheric models.


\section{Method}
\subsection{\Marvel}

The Measured Active RoVibrational Energy Levels (\Marvel) approach
\citep{jt412,07CsCzFu.marvel,12FuCsi.method} is an algorithm that enables
a set of assigned experimental transition frequencies to be converted
into empirical energy levels with associated uncertainties propagated
from the input transition data to the output energy levels.
This conversion relies on the construction of experimental
spectroscopic networks (SNs)
\citep{11CsFuxx.marvel,12FuCsxx.marvel,14FuArMe.marvel,16ArPeFu.marvel}
which contains all inter-connected transitions.  For a detailed
description of the approach, algorithm and program, we refer readers
to \citet{12FuCsi.method}.

The \Marvel\ approach has been used to compile empirical energy levels for
the very important and electronically-similar species
$^{48}$Ti$^{16}$O \citep{jt672}. Other \Marvel\ studies 
on astronomically important molecules include those for
$^{12}$C$_2$ \citep{jt637}, acetylene \citep{jt705}, ammonia
\citep{jt608,jtNH3update}, SO$_2$ \citep{jt704}, H$_2$S \citep{jt718}
and isotopologues of H$_3^+$ \citep{13FuSzMa.marvel,13FuSzFa.marvel}.
These are in addition to energies for the isotopologues of water \citep{jt454,jt482,jt539,jt576}
for which the MARVEL procedure was originally developed \citep{jt562}.

This paper utilised the \Marvel{} algorithm through a specially designed web-interface, available at \url{http://kkrk.chem.elte.hu/marvelonline} \citep{MARVELonline}, making it highly accessible across computer systems without installation of specialised code. Numerous updates to the online interface were also made during this project and related projects in order to optimise the speed, ease and quality of data processing; 
for example, options were made available to automatically update uncertainties within thresholds when processing initial data to find a self-consistent spectroscopic network. 

\begin{figure}
\includegraphics[width=0.5\textwidth]{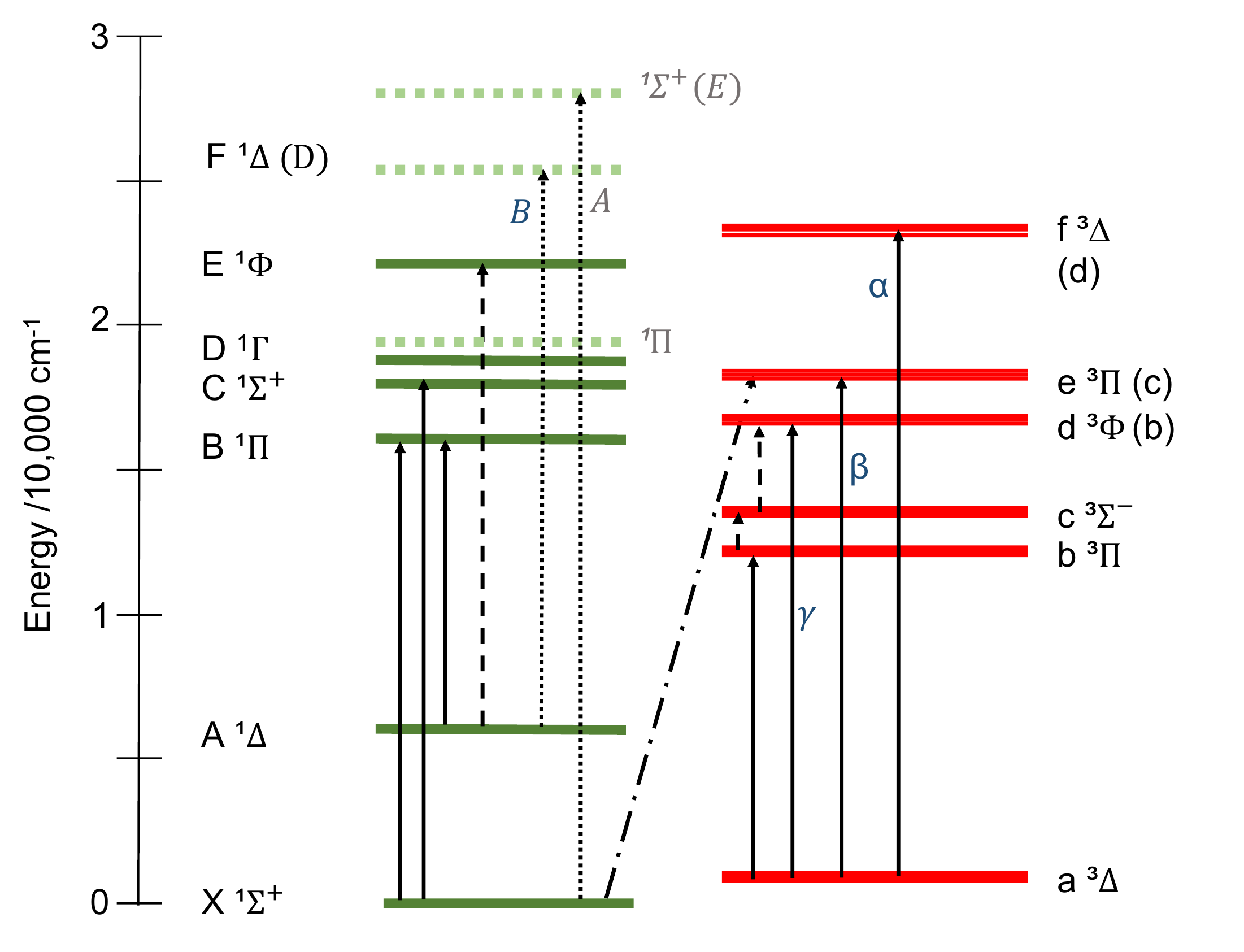}
\caption{\label{fig:fig1} The electronic structure of \ZrO{}, with approximate $T_e$ and labels taken from \cite{90LaBa.ZrO}; where different, labels from \cite{79HuHe} are given in brackets. The solid horizontal lines are those electronic states whose existence and assignment is reasonably secure with reliable theoretical predictions, while the dashed horizontal lines indicate states that some authors have proposed for ZrO but which are not supported by theory or rotationally resolved experiment.  This diagram also shows the main band systems of ZrO, with solid lines showing the bands for which rotationally-resolved allowed transitions have been analysed, the alternating dotted-dashed line representing an experimentally-observed inter-combination band while long dashed lines represent allowed transitions that have not been measured in rotationally-resolved spectra and the short dashed lines represent transitions that have previously, probably erroneously, been assigned as ZrO bands.}
\end{figure}

\subsection{Electronic structure and spectroscopy of ZrO}
ZrO and TiO share similar features in their electronic structure, as
Zr is directly below Ti on the periodic table. Specifically, both have the same qualitative ordering of many low-lying electronic states (in terms of symmetry and spin), with slight differences in $T_e$ so that, e.g. unlike in TiO, the ground electronic
state of ZrO is a spin singlet, \X. Those states with
well-characterised experimental electronic states below 25,000 \cm{}
are shown in \Cref{fig:fig1}, which also gives the observed bands
linking these states. Note that we did not find any
rotationally-resolved spectral data involving the \D{}, \E{}, \Tc{} or
\Tf{} states.

The singlet \X{} ground state has allowed excitations to the \B{} and \C{} states. Significant absorption  also occurs from thermal population of the \Ta{} states to the higher singlet states \Tb{}, \Td{}, \Te{} and \Tf{}. In the high temperature gaseous environments where \ZrO\ is present astrophysically, transitions from the \A{} state to the \B{} and \E{} states may also be relevant. 
 
 \vspace{1em}

\subsection{Quantum numbers and selection rules}
The most obvious information to include in the label of a rovibronic state of ZrO is the
electronic state, $state$, the total angular momentum,  $J$, and the
vibrational quantum number, $v$. We find these to be relatively
unambiguous to define. 

For the triplet states, we also need to provide information about the
electronic spin state; in this case, we choose to include this as part
of the label for the electronic state. The parity of energy levels
usually only influences the energy in a measurable manner for $\Pi$
states; we absorb the $e$ and $f$ parity labels
\citep{75BrHoHu.diatom} into the electronic state label to
reduce the overall number of labels.  





\subsection{Literature Review}

In the first half of the  twentieth century, there was considerable interest in 
studying the visible and ultraviolet spectrum of \ZrO, with many bandheads 
measured by \cite{32Lowater.ZrO}, \cite{49Herbig.ZrO}, \cite{49Afaf.ZrO}, 
\cite{50Afaf.ZrO} and \cite{50Afaf2.ZrO}. These studies include many involving
transitions to electronic 
states that have yet to  be  investigated using rotationally-resolved 
spectra.

More recently, there was an extensive experimental effort over the 1970s to 
early 1980s by various groups to obtain rotationally resolved assigned 
experimental spectra for various important \ZrO{} bands; these studies as well 
as more recent rotationally resolved studies are summarized in  
\Cref{tab:datasources}. 

Two further studies in the 1980s, \citet{81HaDaZo.ZrO} and \cite{88StMoKu.ZrO}, 
investigated higher vibrational levels of some of the most important electronic states but without rotational resolution. 






\begin{center}
\begin{longtable*}{llcccrcllllllll}
\label{tab:datasources} \\
\multicolumn{8}{c}{{ \tablename\ \thetable{} -- \textsc{Data sources and their characteristics for \ZrO. A/V means Available/Validated.}}} 
\\
\toprule
Tag &  Ref & Band & & Range(\cm{}) &  $J$ Range  & Trans. (A/V) & \mc{3}{c}{Uncertainties (\cm{})} \\
\cline{8-10}
& & &  &  & & &    Min & Av & Max \\
\midrule
\endfirsthead

\multicolumn{5}{c}%
{{ \tablename\ \thetable{} -- continued from previous page}} \\
\toprule

Tag &  Ref & Band & & Range(\cm{}) &  $J$  Range  & Trans.(A/V) & \mc{3}{c}{Uncertainties (\cm{})} \\
\cline{8-10}
& & &  &  & & &    Min & Av & Max \\
\midrule
\endhead

\bottomrule
\multicolumn{5}{c}{{Continued on next page}} \\
\endfoot

\bottomrule
\endlastfoot
\vspace{-0.5em}\\
54LaUhBa & \cite{54UhlerArkiv281.ZrO} & \Td$_{2}$ \enspace-- \Ta$_{1}$ & 0 - 0 &   15282 - 15442 & 11 - \enspace89 &159/159 &0.1 & 0.1 & 0.28\\ 
& & \Td$_{3}$ \enspace-- \Ta$_{2}$ & 0 - 0 &   15612 - 15755 & 11 - \enspace93 &149/149 &0.1 & 0.1 & 0.32\\ 
& & \Td$_{4}$ \enspace-- \Ta$_{3}$ & 0 - 0 &   15898 - 16048 & 11 - \enspace95 &165/165 &0.1 & 0.11 & 0.34\\ 
& & \Tf$_{1}$ \enspace-- \Ta$_{1}$ & 0 - 0 &   21351 - 21542 & 20 - \enspace76 &105/105 &0.1 & 0.1 & 0.31\\ 
& & \Tf$_{2}$ \enspace-- \Ta$_{2}$ & 0 - 0 &   21351 - 21555 & 20 - \enspace80 &111/111 &0.1 & 0.1 & 0.1\\ 
& & \Tf$_{3}$ \enspace-- \Ta$_{3}$ & 0 - 0 &   21457 - 21640 & 20 - \enspace81 &106/106 &0.1 & 0.1 & 0.1\\ 
\vspace{-0.5em}\\ 
54Uhler & \cite{54UhlerArkiv295.ZrO} & \Te$_{0e}$ \enspace-- \Ta$_{1}$ & 0 - 0 &   17884 - 18002 & 15 - \enspace60 &106/106 &0.1 & 0.1 & 0.14\\ 
& & \Te$_{0f}$ \enspace-- \Ta$_{1}$ & 0 - 0 &   17889 - 18006 & 27 - \enspace59 &96/91 &0.1 & 0.1 & 0.3\\ 
& & \Te$_{1e}$ \enspace-- \Ta$_{2}$ & 0 - 0 &   17619 - 17757 & 20 - \enspace74 &103/91 &0.1 & 0.12 & 0.31\\ 
& & \Te$_{1f}$ \enspace-- \Ta$_{2}$ & 0 - 0 &   17653 - 17758 & 19 - \enspace59 &99/63 &0.1 & 0.19 & 0.62\\ 
& & \Te$_{2}$ \enspace-- \Ta$_{3}$ & 0 - 0 &   17326 - 17483 & 13 - \enspace85 &143/139 &0.1 & 0.15 & 0.43\\ 
\vspace{-0.5em}\\ 
57Akerlind & \cite{57Akerlind.ZrO} & \F \enspace-- \A & 0 - 0 &   18994 - 19280 & 17 - 102 &156/156 &0.1 & 0.1 & 0.11\\ 
& & \F \enspace-- \A & 1 - 0 &   19843 - 20106 & 35 - \enspace94 &110/110 &0.1 & 0.1 & 0.15\\ 
\vspace{-0.5em}\\ 
73BaTa & \cite{73BaTa.ZrO} & \B$_e$ \enspace-- \X & 0 - 0 &   15136 - 15391 & 18 - 107 &149/145 &0.01 & 0.031 & 0.16\\ 
& & \B$_f$ \enspace-- \X & 0 - 0 &   15185 - 15382 & 8 - \enspace96 &85/83 &0.01 & 0.024 & 0.13\\ 
\vspace{-0.5em}\\ 
73Lindgren & \cite{73Lindgren.ZrO} & \Te$_{1e}$ \enspace-- \Ta$_{1}$ & 0 - 0 &   17995 - 18050 & 30 - \enspace61 &53/53 &0.07 & 0.086 & 0.19\\ 
& & \Te$_{1f}$ \enspace-- \Ta$_{1}$ & 0 - 0 &   17991 - 18048 & 30 - \enspace60 &51/50 &0.07 & 0.09 & 0.23\\ 
& & \Te$_{2}$ \enspace-- \Ta$_{2}$ & 0 - 0 &   17761 - 17820 & 47 - \enspace65 &36/29 &0.07 & 0.12 & 0.35\\ 
\vspace{-0.5em}\\ 
76PhDa.CX & \cite{76PhDaCX.ZrO} & \C \enspace-- \X & 0 - 0 &   16732 - 17060 & 2 - 121 &232/203 &0.02 & 0.044 & 0.19\\ 
\vspace{-0.5em}\\ 
76PhDa.BX & \cite{76PhDaBX.ZrO} & \B$_e$ \enspace-- \X & 0 - 0 &   15102 - 15391 & 5 - 132 &201/188 &0.02 & 0.039 & 0.14\\ 
& & \B$_e$ \enspace-- \X & 0 - 1 &   14292 - 14423 & 1 - 102 &144/135 &0.02 & 0.046 & 0.18\\ 
& & \B$_e$ \enspace-- \X & 0 - 2 &   13244 - 13431 & 17 - 116 &101/100 &0.02 & 0.042 & 0.1\\ 
& & \B$_e$ \enspace-- \X & 1 - 0 &   16023 - 16244 & 1 - 107 &149/148 &0.02 & 0.046 & 0.16\\ 
& & \B$_e$ \enspace-- \X & 1 - 2 &   14038 - 14313 & 4 - 116 &177/175 &0.02 & 0.045 & 0.31\\ 
& & \B$_e$ \enspace-- \X & 1 - 3 &   13246 - 13359 & 1 - 102 &135/134 &0.02 & 0.042 & 0.15\\ 
& & \B$_e$ \enspace-- \X & 2 - 0 &   16817 - 17091 & 4 - 117 &159/142 &0.02 & 0.056 & 0.2\\ 
& & \B$_e$ \enspace-- \X & 2 - 1 &   15936 - 16122 & 1 - 104 &146/136 &0.02 & 0.046 & 0.27\\ 
& & \B$_e$ \enspace-- \X & 2 - 3 &   14046 - 14205 & 1 - 106 &136/135 &0.02 & 0.048 & 0.17\\ 
& & \B$_e$ \enspace-- \X & 2 - 4 &   13122 - 13257 & 2 - 108 &150/147 &0.02 & 0.034 & 0.14\\ 
& & \B$_e$ \enspace-- \X & 3 - 1 &   16690 - 16963 & 1 - \enspace90 &114/111 &0.02 & 0.038 & 0.14\\ 
& & \B$_e$ \enspace-- \X & 3 - 5 &   13051 - 13157 & 1 - 100 &135/132 &0.02 & 0.036 & 0.24\\ 
& & \B$_e$ \enspace-- \X & 3 - 6 &   11990 - 12223 & 1 - 136 &188/185 &0.02 & 0.036 & 0.15\\ 
& & \B$_e$ \enspace-- \X & 4 - 2 &   16547 - 16836 & 2 - 116 &139/139 &0.02 & 0.04 & 0.3\\ 
& & \B$_e$ \enspace-- \X & 4 - 5 &   13746 - 13991 & 1 - 104 &102/100 &0.02 & 0.043 & 0.3\\ 
& & \B$_e$ \enspace-- \X & 4 - 6 &   12824 - 13057 & 2 - 108 &132/132 &0.02 & 0.042 & 0.22\\ 
& & \B$_e$ \enspace-- \X & 5 - 3 &   16659 - 16710 & 2 - \enspace61 &46/46 &0.02 & 0.029 & 0.064\\ 
& & \B$_e$ \enspace-- \X & 5 - 7 &   12822 - 12959 & 2 - 108 &130/130 &0.02 & 0.026 & 0.13\\ 
& & \B$_f$ \enspace-- \X & 0 - 0 &   15108 - 15383 & 1 - 113 &114/109 &0.02 & 0.036 & 0.17\\ 
& & \B$_f$ \enspace-- \X & 0 - 1 &   14289 - 14414 & 2 - \enspace80 &77/73 &0.02 & 0.047 & 0.44\\ 
& & \B$_f$ \enspace-- \X & 0 - 2 &   13250 - 13431 & 34 - 107 &70/70 &0.02 & 0.034 & 0.14\\ 
& & \B$_f$ \enspace-- \X & 1 - 0 &   16021 - 16237 & 1 - \enspace96 &93/93 &0.02 & 0.037 & 0.1\\ 
& & \B$_f$ \enspace-- \X & 1 - 3 &   13248 - 13349 & 3 - \enspace76 &72/72 &0.02 & 0.036 & 0.16\\ 
& & \B$_f$ \enspace-- \X & 2 - 0 &   16759 - 17084 & 2 - 113 &96/94 &0.02 & 0.046 & 0.2\\ 
& & \B$_f$ \enspace-- \X & 2 - 1 &   15938 - 16115 & 3 - \enspace87 &81/76 &0.02 & 0.04 & 0.19\\ 
& & \B$_f$ \enspace-- \X & 2 - 3 &   14039 - 14195 & 9 - \enspace90 &72/72 &0.02 & 0.04 & 0.15\\ 
& & \B$_f$ \enspace-- \X & 2 - 4 &   13122 - 13248 & 1 - \enspace85 &80/77 &0.02 & 0.033 & 0.097\\ 
& & \B$_f$ \enspace-- \X & 3 - 1 &   16703 - 16957 & 1 - 100 &90/87 &0.02 & 0.04 & 0.19\\ 
& & \B$_f$ \enspace-- \X & 3 - 5 &   13050 - 13148 & 2 - \enspace75 &69/68 &0.02 & 0.035 & 0.2\\ 
& & \B$_f$ \enspace-- \X & 3 - 6 &   11986 - 12213 & 3 - 121 &99/97 &0.02 & 0.033 & 0.11\\ 
& & \B$_f$ \enspace-- \X & 4 - 2 &   16555 - 16830 & 1 - 104 &81/79 &0.02 & 0.037 & 0.19\\ 
& & \B$_f$ \enspace-- \X & 4 - 5 &   13748 - 13983 & 1 - 110 &74/73 &0.02 & 0.033 & 0.19\\ 
& & \B$_f$ \enspace-- \X & 4 - 6 &   12834 - 13047 & 8 - 111 &96/96 &0.02 & 0.035 & 0.19\\ 
& & \B$_f$ \enspace-- \X & 5 - 3 &   16661 - 16704 & 3 - \enspace41 &30/30 &0.02 & 0.025 & 0.06\\ 
& & \B$_f$ \enspace-- \X & 5 - 7 &   12822 - 12950 & 1 - \enspace86 &77/77 &0.02 & 0.022 & 0.06\\ 
\vspace{-0.5em}\\ 
79GaDe & \cite{79GaDe.ZrO} & \X \enspace-- \X & 1 - 0 &     952 - \enspace986 & 1 - \enspace20 &40/33 &0.02 & 0.075 & 0.42\\ 
\vspace{-0.5em}\\ 
79PhDa & \cite{79PhDa.ZrO} & \Td$_{2}$ \enspace-- \Ta$_{1}$ & 0 - 0 &   15132 - 15442 & 2 - 150 &372/371 &0.02 & 0.037 & 0.14\\ 
& & \Td$_{2}$ \enspace-- \Ta$_{1}$ & 0 - 1 &   14172 - 14515 & 2 - 150 &351/350 &0.02 & 0.042 & 0.13\\ 
& & \Td$_{2}$ \enspace-- \Ta$_{1}$ & 1 - 0 &   15862 - 16289 & 2 - 151 &393/393 &0.02 & 0.04 & 0.14\\ 
& & \Td$_{2}$ \enspace-- \Ta$_{1}$ & 1 - 1 &   15089 - 15361 & 2 - 150 &327/318 &0.02 & 0.046 & 0.2\\ 
& & \Td$_{2}$ \enspace-- \Ta$_{1}$ & 1 - 2 &   14093 - 14440 & 2 - 151 &358/357 &0.02 & 0.041 & 0.16\\ 
& & \Td$_{2}$ \enspace-- \Ta$_{1}$ & 2 - 1 &   15814 - 16191 & 3 - 151 &356/355 &0.02 & 0.047 & 0.17\\ 
& & \Td$_{2}$ \enspace-- \Ta$_{1}$ & 2 - 2 &   14928 - 15277 & 1 - 149 &327/314 &0.02 & 0.046 & 0.2\\ 
& & \Td$_{2}$ \enspace-- \Ta$_{1}$ & 2 - 3 &   14015 - 14364 & 2 - 151 &363/362 &0.02 & 0.045 & 0.17\\ 
& & \Td$_{2}$ \enspace-- \Ta$_{1}$ & 3 - 2 &   15679 - 16113 & 2 - 151 &411/407 &0.02 & 0.049 & 0.2\\ 
& & \Td$_{2}$ \enspace-- \Ta$_{1}$ & 3 - 3 &   14929 - 15194 & 2 - 137 &332/329 &0.02 & 0.039 & 0.12\\ 
& & \Td$_{2}$ \enspace-- \Ta$_{1}$ & 3 - 4 &   13944 - 14289 & 3 - 151 &371/370 &0.02 & 0.038 & 0.17\\ 
& & \Td$_{2}$ \enspace-- \Ta$_{1}$ & 3 - 5 &   13214 - 13384 & 2 - 101 &206/206 &0.02 & 0.051 & 0.14\\ 
& & \Td$_{2}$ \enspace-- \Ta$_{1}$ & 4 - 3 &   15802 - 16024 & 2 - 101 &261/261 &0.02 & 0.041 & 0.13\\ 
& & \Td$_{2}$ \enspace-- \Ta$_{1}$ & 4 - 5 &   13877 - 14214 & 2 - 151 &348/348 &0.02 & 0.039 & 0.18\\ 
& & \Td$_{2}$ \enspace-- \Ta$_{1}$ & 5 - 6 &   13947 - 14136 & 2 - 101 &209/209 &0.02 & 0.023 & 0.13\\ 
& & \Td$_{3}$ \enspace-- \Ta$_{2}$ & 0 - 0 &   15417 - 15754 & 2 - 144 &375/361 &0.02 & 0.033 & 0.2\\ 
& & \Td$_{3}$ \enspace-- \Ta$_{2}$ & 0 - 1 &   14499 - 14675 & 83 - 150 &83/77 &0.02 & 0.052 & 0.18\\ 
& & \Td$_{3}$ \enspace-- \Ta$_{2}$ & 1 - 0 &   16163 - 16602 & 2 - 150 &397/395 &0.02 & 0.036 & 0.53\\ 
& & \Td$_{3}$ \enspace-- \Ta$_{2}$ & 1 - 1 &   15423 - 15665 & 1 - 133 &307/291 &0.02 & 0.037 & 0.17\\ 
& & \Td$_{3}$ \enspace-- \Ta$_{2}$ & 1 - 2 &   14490 - 14749 & 2 - 151 &363/358 &0.02 & 0.037 & 0.2\\ 
& & \Td$_{3}$ \enspace-- \Ta$_{2}$ & 2 - 1 &   16065 - 16514 & 2 - 151 &375/361 &0.02 & 0.037 & 0.29\\ 
& & \Td$_{3}$ \enspace-- \Ta$_{2}$ & 2 - 2 &   15198 - 15591 & 1 - 147 &374/353 &0.02 & 0.049 & 0.2\\ 
& & \Td$_{3}$ \enspace-- \Ta$_{2}$ & 2 - 3 &   14309 - 14673 & 2 - 151 &365/336 &0.02 & 0.041 & 0.17\\ 
& & \Td$_{3}$ \enspace-- \Ta$_{2}$ & 3 - 2 &   15972 - 16422 & 2 - 151 &387/353 &0.02 & 0.047 & 0.2\\ 
& & \Td$_{3}$ \enspace-- \Ta$_{2}$ & 3 - 3 &   15102 - 15508 & 2 - 151 &358/333 &0.02 & 0.053 & 0.24\\ 
& & \Td$_{3}$ \enspace-- \Ta$_{2}$ & 3 - 4 &   14230 - 14594 & 3 - 150 &308/303 &0.02 & 0.031 & 0.19\\ 
& & \Td$_{3}$ \enspace-- \Ta$_{2}$ & 3 - 5 &   13516 - 13696 & 2 - 101 &178/177 &0.02 & 0.043 & 0.15\\ 
& & \Td$_{3}$ \enspace-- \Ta$_{2}$ & 4 - 3 &   16103 - 16335 & 2 - 101 &244/244 &0.02 & 0.037 & 0.13\\ 
& & \Td$_{3}$ \enspace-- \Ta$_{2}$ & 4 - 5 &   14150 - 14518 & 3 - 151 &336/336 &0.02 & 0.038 & 0.2\\ 
& & \Td$_{3}$ \enspace-- \Ta$_{2}$ & 5 - 6 &   14250 - 14445 & 2 - 101 &222/222 &0.02 & 0.024 & 0.081\\ 
& & \Td$_{4}$ \enspace-- \Ta$_{3}$ & 0 - 0 &   15704 - 16040 & 3 - 151 &360/330 &0.02 & 0.053 & 0.23\\ 
& & \Td$_{4}$ \enspace-- \Ta$_{3}$ & 0 - 1 &   14833 - 15117 & 3 - 151 &370/350 &0.02 & 0.04 & 0.19\\ 
& & \Td$_{4}$ \enspace-- \Ta$_{3}$ & 1 - 0 &   16488 - 16898 & 3 - 147 &363/348 &0.02 & 0.045 & 0.24\\ 
& & \Td$_{4}$ \enspace-- \Ta$_{3}$ & 1 - 1 &   15666 - 15967 & 3 - 150 &336/321 &0.02 & 0.053 & 0.2\\ 
& & \Td$_{4}$ \enspace-- \Ta$_{3}$ & 1 - 2 &   14686 - 15042 & 4 - 151 &395/382 &0.02 & 0.042 & 0.17\\ 
& & \Td$_{4}$ \enspace-- \Ta$_{3}$ & 2 - 1 &   16358 - 16809 & 3 - 151 &403/397 &0.02 & 0.044 & 0.2\\ 
& & \Td$_{4}$ \enspace-- \Ta$_{3}$ & 2 - 2 &   15604 - 15885 & 3 - 147 &311/298 &0.02 & 0.059 & 0.2\\ 
& & \Td$_{4}$ \enspace-- \Ta$_{3}$ & 2 - 3 &   14667 - 14970 & 3 - 136 &372/365 &0.02 & 0.038 & 0.18\\ 
& & \Td$_{4}$ \enspace-- \Ta$_{3}$ & 3 - 2 &   16270 - 16724 & 3 - 151 &393/386 &0.02 & 0.045 & 0.2\\ 
& & \Td$_{4}$ \enspace-- \Ta$_{3}$ & 3 - 3 &   15566 - 15796 & 3 - 108 &239/237 &0.02 & 0.049 & 0.17\\ 
& & \Td$_{4}$ \enspace-- \Ta$_{3}$ & 3 - 4 &   14685 - 14897 & 3 - 105 &262/262 &0.02 & 0.034 & 0.18\\ 
& & \Td$_{4}$ \enspace-- \Ta$_{3}$ & 3 - 5 &   13813 - 13991 & 3 - 101 &187/186 &0.02 & 0.045 & 0.15\\ 
& & \Td$_{4}$ \enspace-- \Ta$_{3}$ & 4 - 3 &   16410 - 16636 & 3 - 101 &251/250 &0.02 & 0.045 & 0.16\\ 
& & \Td$_{4}$ \enspace-- \Ta$_{3}$ & 4 - 5 &   14469 - 14824 & 3 - 151 &385/385 &0.02 & 0.031 & 0.16\\ 
& & \Td$_{4}$ \enspace-- \Ta$_{3}$ & 5 - 6 &   14558 - 14752 & 3 - 101 &224/224 &0.02 & 0.023 & 0.11\\ 
\vspace{-0.5em}\\ 
80HaDa & \cite{80HaDa.ZrO} & \Te$_{1e}$ \enspace-- \X$_{}$ & 0 - 0 &   19047 - 19121 & 29 - \enspace41 &24/23 &0.01 & 0.025 & 0.085\\ 
& & \Te$_{1e}$ \enspace-- \X$_{}$ & 0 - 1 &   18094 - 18152 & 32 - \enspace39 &8/6 &0.01 & 0.032 & 0.071\\ 
& & \Te$_{1e}$ \enspace-- \Ta$_{2}$ & 0 - 0 &   17693 - 17735 & 30 - \enspace41 &22/21 &0.01 & 0.024 & 0.097\\ 
& & \Te$_{1f}$ \enspace-- \X$_{}$ & 0 - 0 &   19085 - 19095 & 29 - \enspace37 &9/9 &0.01 & 0.022 & 0.042\\ 
& & \Te$_{1f}$ \enspace-- \X$_{}$ & 0 - 1 &   18120 - 18128 & 29 - \enspace35 &7/6 &0.01 & 0.012 & 0.015\\ 
& & \Te$_{1f}$ \enspace-- \Ta$_{2}$ & 0 - 0 &   17695 - 17736 & 26 - \enspace41 &28/26 &0.01 & 0.022 & 0.081\\ 
\vspace{-0.5em}\\ 
81HaDa & \cite{81HaDa.ZrO} & \B$_e$ \enspace-- \A & 0 - 0 &    9102 - \enspace9507 & 2 - 147 &394/354 &0.01 & 0.017 & 0.14\\ 
& & \B$_e$ \enspace-- \A & 1 - 0 &   10057 - 10359 & 3 - 109 &176/176 &0.01 & 0.024 & 0.13\\ 
& & \B$_e$ \enspace-- \A & 2 - 1 &   10022 - 10271 & 3 - 115 &151/149 &0.01 & 0.023 & 0.17\\ 
& & \B$_f$ \enspace-- \A & 0 - 0 &    9152 - \enspace9507 & 2 - 149 &373/336 &0.01 & 0.016 & 0.09\\ 
& & \B$_f$ \enspace-- \A & 1 - 0 &   10110 - 10359 & 3 - 118 &179/178 &0.01 & 0.021 & 0.12\\ 
& & \B$_f$ \enspace-- \A & 2 - 1 &   10033 - 10271 & 3 - 115 &159/155 &0.01 & 0.022 & 0.13\\ 
\vspace{-0.5em}\\ 
81HaDaZo & \cite{81HaDaZo.ZrO} & \B$_e$ \enspace-- \A & 1 - 0 &   10324 - 10359 & 14 - \enspace22 &21/21 &0.01 & 0.011 & 0.02\\ 
& & \B$_e$ \enspace-- \Ta & 1 - 1 &   13915 - 13948 & 15 - \enspace21 &11/11 &0.01 & 0.01 & 0.01\\ 
\vspace{-0.5em}\\ 
88SiMiHuHa & \cite{88SiMiHu.ZrO} & \C \enspace-- \X & 0 - 0 &   17011 - 17060 & 0 - \enspace30 &59/57 &0.006 & 0.025 & 0.1\\ 
\vspace{-0.5em}\\ 

90SuLoFrMa & \cite{90SuLoFrMa.ZrO} & \X \enspace-- \X & 0 - 0 &       0 - \enspace\enspace\enspace\enspace1 & 0 - \enspace\enspace1 &1/1 &3$\times 10^{-7}$ & 3$\times 10^{-7}$ & 3$\times 10^{-7}$\\ 
\vspace{-0.5em}\\ 
94Jonsson & \cite{94Jonsson.ZrO} & \Tb$_{0e}$ \enspace-- \Ta$_{1}$ & 0 - 0 &   10535 - 10714 & 7 - 107 &180/180 &0.006 & 0.012 & 0.069\\ 
& & \Tb$_{0f}$ \enspace-- \Ta$_{1}$ & 0 - 0 &   10579 - 10702 & 14 - \enspace92 &113/113 &0.006 & 0.0078 & 0.026\\ 
& & \Tb$_{1e}$ \enspace-- \Ta$_{2}$ & 0 - 0 &   10611 - 10728 & 11 - 100 &137/137 &0.006 & 0.0088 & 0.045\\ 
& & \Tb$_{1f}$ \enspace-- \Ta$_{2}$ & 0 - 0 &   10625 - 10731 & 20 - \enspace90 &104/104 &0.006 & 0.0097 & 0.067\\ 
& & \Tb$_{2e}$ \enspace-- \Ta$_{3}$ & 0 - 0 &   10614 - 10750 & 11 - 111 &171/171 &0.006 & 0.0075 & 0.049\\ 
& & \Tb$_{2f}$ \enspace-- \Ta$_{3}$ & 0 - 0 &   10614 - 10750 & 11 - 111 &168/168 &0.006 & 0.0079 & 0.049\\ 
\vspace{-0.5em}\\ 
95KaMcHe & \cite{95KaMcHe.ZrO} & \Te$_{1e}$ \enspace-- \Ta$_{1}$ & 0 - 0 &   17993 - 18050 & 2 - \enspace63 &67/66 &0.007 & 0.013 & 0.057\\ 
& & \Te$_{1f}$ \enspace-- \Ta$_{1}$ & 0 - 0 &   17984 - 18048 & 2 - \enspace67 &79/79 &0.007 & 0.016 & 0.084\\ 
& & \Te$_{2}$ \enspace-- \Ta$_{2}$ & 0 - 0 &   17748 - 17820 & 2 - \enspace64 &95/92 &0.007 & 0.021 & 0.31\\ 
\vspace{-0.5em}\\ 
\vspace{-0.5em}\\ 
99BeGe & \cite{99BeGe.ZrO} & \X \enspace-- \X & 0 - 0 &       1 - \enspace\enspace\enspace\enspace1 & 0 - \enspace\enspace1 &1/1 &1$\times 10^{-7}$ & 1$\times 10^{-7}$ & 1$\times 10^{-7}$\\ 
& & \X \enspace-- \X & 1 - 1 &       0 - \enspace\enspace\enspace\enspace1 & 0 - \enspace\enspace1 &1/1 &1$\times 10^{-7}$ & 1$\times 10^{-7}$ & 1$\times 10^{-7}$\\ 
& & \X \enspace-- \X & 2 - 2 &       0 - \enspace\enspace\enspace\enspace1 & 0 - \enspace\enspace1 &1/1 &1$\times 10^{-7}$ & 1$\times 10^{-7}$ & 1$\times 10^{-7}$\\ 
& & \X \enspace-- \X & 3 - 3 &       0 - \enspace\enspace\enspace\enspace1 & 0 - \enspace\enspace1 &1/1 &1$\times 10^{-7}$ & 1$\times 10^{-7}$ & 1$\times 10^{-7}$\\ 
\end{longtable*}
\end{center}

\begin{table*}
\footnotesize
 
\centering
\caption{\label{tab:otherexp} Experimental ZrO papers not  used in the rotationally-resolved \Marvel\ or bandhead analysis. }
\begin{tabular}{Hlp{12cm}}
\toprule
Tag  &  Ref  &  Comment   \\
\midrule
35Lowater & \citet{35Lowater.ZrO} & Good analysis of vibronic bands (including 
triplet splitting), but the rotational analysis of the \Tf -- \Ta{} is 
incorrect.\\ 
41TaHo & \citet{41TaHo.ZrO} & Incorrect rotational analysis of the 
\Tb -- \Ta{} bands and more recent data are available \\
48Kiess	 &  \citet{48Kiess.ZrO}  &  Data not available online. \\
49Herbig  &  \citet{49Herbig.ZrO}  &  Unassigned.	\\
 &  \citet{54Uhler.ZrO}  &  Rotational analysis with band constants, but 
assigned line positions are given in the associated papers 
\citep{54UhlerArkiv281.ZrO,54UhlerArkiv295.ZrO}. \\
& \citet{55UlAA.ZrO} & Rotational analysis with band constants for singlet A 
system, but assigned line positions are given in the associated paper 
\citep{56UhAk.ZrO} \\
& \citet{56AAker.ZrO} &Band constants from analysis of a system assigned as the 
singlet B system at 8192~\AA, but this is not consistent with 
\cite{57Akerlind.ZrO}. \\
& \citet{57Akerlind.ZrO} & Rotationally-resolved data from a system assigned as 
the singlet B system, but with frequencies around 19,000 \cm{}, not 12,000 \cm{} as 
indicated by the 8192 \AA{} labelling. Due to this confusion, and some later 
papers \citep{79PhDa.ZrO+,80BaLi.ZrO+}  that provide good evidence that the 8192 
\AA{} band(around 12,000 \cm{}) is a \ce{ZrO+} band, these data are not included 
in our compilation. \\ 
73TaBa  &  \citet{73TaBa.ZrO}  &  Data  very poorly reproduced digitally, and 
higher resolution spectra for the \Td--\Ta{} bands studied are available. 
\\
& \cite{65WeMc.ZrO} & Ground state determination in Ne matrix. \\
74ScSu  &  \citet{74ScSu.ZrO}  &  Complete and systematic analysis of available 
data for triplet systems, but provides no assigned rotationally resolved data.	\\
74BiCmCy  &  \cite{74BiCmCy.ZrO}  &  Determination of Singlet-Triplet 
separation, but no assigned rotational data.	\\
76LaBrBr	 &  \citet{76LaBrBr.ZrO}  &  ZrO in Ne inert matrix at 4K\\
79PhDaGa & \cite{79PhDaGa.ZrO} & The rotational analysis here was shown to be 
incorrect by the subsequent re-analysis by \cite{94Jonsson.ZrO}; see text. \\
79GaDe &  \cite{79GaDe.ZrO} & Rotationally unresolved infrared study; used for 
comparison against bandheads but not as part of the \Marvel{} dataset. \\
80Murty  &  \citet{80Murty.ZrO}  &  Contains molecular constants for \Te{} and 
\Tc, but provides no assigned rotationally resolved data.   \\
&  \citet{81HaDaZo.ZrO}  &  Identification of bands in astronomical vs. 
laboratory spectra, not rotational analysis.  \\
87Afaf &  \citet{87Afaf.ZrO}  &  Reanalysis of data and recommended molecular 
constants; also proposes a singlet C band from \X{} to a singlet at 7870 \cm{} 
above; this was later discounted(e.g. \cite{95Afaf.ZrO}). \\
88DaHa	 &  \citet{88DaHa.ZrO}  &  Consolidation of data and proposed electronic 
structure.\\
88SiMiHeHa & \cite{88SiMiHeHa.ZrO} & High resolution study of the \Te -- \Ta{} 
0-0 band; assigned line positions were not published with the original data and 
could not be located. \\
95Afaf	 &  \citet{95Afaf.ZrO}  &  Discusses the $\delta$($^3\Pi$ -- \Ta{}) and 
$\phi$($^3\Delta$ -- \Ta{}) bands, but provides no assigned rotationally resolved data. \\
10BaCh  &  \citet{10BaCh.ZrO}  &  Low resolution data with bandhead information 
on the \C--\X{} state.  \\ 
\bottomrule
\end{tabular}
\end{table*}

Beyond the wavenumber of the lines, many experimental studies have focused on the intensity of transitions, e.g. \citep{49Herbig.ZrO,80MuPr.ZrO,85LiDa.ZrO,93LiDaSo.ZrO}, radiative lifetimes, e.g. \citep{78Hammer.ZrO,79HaDa.ZrO,88SiMiHu.ZrO} and permanent dipole moments, e.g. \citep{90SuLoFrMa.ZrO,00PeKoLiLu.ZrO}. 

The partition function and dissociation constants for zirconium oxide have been considered by various authors, including \citet{83ShLi.ZrO}.

\begin{table*}
\centering
\caption{\label{tab:marvelinput}
Extract from the 90Zr-16O.marvel.inp input file for \ZrO.}
\footnotesize \tabcolsep=5pt
\begin{tabular}{lllcclccr}
\\
\toprule
  \mc{1}{c}{1}      &         \mc{1}{c}{2}         &    \mc{1}{c}{3}   &    \mc{1}{c}{4}   &   \mc{1}{c}{5}   &   \mc{1}{c}{6}   &    \mc{1}{c}{7}   &   \mc{1}{c}{8}   &   \mc{1}{c}{9}  \\
    \midrule
    \multicolumn{1}{c}{$\tilde{\nu}$}  &  \multicolumn{1}{c}{$\Delta\tilde{\nu}$}  &   \mc{1}{c}{State$^\prime$}     &  \multicolumn{1}{c}{$v^\prime$}  &   $J^\prime$   &     \mc{1}{c}{State$^{\prime\prime}$}    &   \multicolumn{1}{c}{$v^{\prime\prime}$}  &  $J^{\prime\prime}$    &     \mc{1}{c}{ID}  \\
     \midrule

17059.5189  & 	0.006000 & 	C1Sigma+  & 0  & 18  & 		X1Sigma+  & 0  & 17 	 & 	88SiMiHuHa.46 \\
17059.9101 & 	0.006000 & 	C1Sigma+  & 0  & 21  & 		X1Sigma+ &  0  & 20  & 		88SiMiHuHa.49\\
17059.9295 & 	0.006000 & 	C1Sigma+  & 0  & 25  & 		X1Sigma+ &  0  & 24  & 		88SiMiHuHa.53\\
17059.9792 & 	0.006000 & 	C1Sigma+  & 0 &  24  & 		X1Sigma+ &  0  & 23  & 		88SiMiHuHa.52\\
10710.4902 & 	0.006622 & 	b3Pi\_2e &  0  & 46 	 & 	a3Delta\_3  & 0 &  46 	 & 	94Jonsson.540\\
10710.4902 & 	0.006622 & 	b3Pi\_2f &  0  & 46 	 & 	a3Delta\_3  & 0  & 46 	 & 	94Jonsson.730\\
10617.7781 & 	0.006660 & 	b3Pi\_2e &  0  & 79  & 		a3Delta\_3 &  0  & 80  & 		94Jonsson.690\\
\bottomrule
\end{tabular}

\begin{tabular}{ccl}
\\
             Column        &     Notation                  &       \\
\midrule
   1  &    $\tilde{\nu}$              &     Transition frequency(in \cm) \\
   2  &  $\Delta\tilde{\nu}$         &    Estimated uncertainty in transition frequency(in \cm) \\
   3  &   State$^\prime$  &   Electronic state of upper energy level; also includes parity for $\Pi$ states and $\Omega$ for triplet states \\
   4  &  $v^\prime$  &  Vibrational quantum number  of upper  level \\ 
   5  &   $J^\prime$                &        Total angular momentum of upper  level   \\
   6  &   State$^{\prime\prime}$  &   Electronic state of lower energy level; also includes parity for $\Pi$ states and $\Omega$ for triplet states \\
   7  &  $v^{\prime\prime}$  &  Vibrational quantum number  of lower  level \\
   8  &   $J^{\prime\prime}$                &        Total angular momentum of lower  level   \\
   9  &  ID  &  Unique ID for transition, with reference key for source(see \Cref{tab:datasources}) and counting number \\
\bottomrule
\end{tabular}
\end{table*}

There are a number of other studies of ZrO spectra which we have not used in this study for various 
reasons. These data sources are collated in \Cref{tab:otherexp} with brief comments. 

The data in \cite{73TaBa.ZrO} was of very poor readability, which meant the 
accuracy of digitisation even manually could not be guaranteed. As there are
substantial more modern data available for the same transitions, we did not use 
these data.

A key omission to our \Marvel{} collation is the \cite{79PhDaGa.ZrO} paper; the 
subsequent study by \cite{94Jonsson.ZrO} performed a complete re-analysis of the 
ZrO spectra in the region around 10,750 \cm{} that assigned all bands whereas 
the \cite{79PhDaGa.ZrO} analysis omitted many bands. A key feature of the 
\cite{94Jonsson.ZrO} analysis was a large $\Lambda$-doubling splitting between the 
\Tb$_{0e}$ and the \Tb$_{0f}$ state. This is attributed to spin-orbit 
coupling with the nearby \Tc{} state whose $T_e$ was only predicted semi-quantitatively with 
computational techniques in the 1990s. 

Unfortunately, the data of \cite{88SiMiHeHa.ZrO}  could not be located; however, the 
spectra and analysis by \cite{95KaMcHe.ZrO} covers the same spectral 
transitions. 

Finally, we want to briefly discuss 
\cite{10BaCh.ZrO} in more detail, particularly their claim to observe a $^1\Pi$ 
-- \X{} system near 19,480 \cm{}. 
 We strongly question this assignment because there is no expectation of a 
$^1\Pi$ state in this energy range from either \emph{ab initio} calculations or analogy to the TiO electronic states.  Based on vibrational frequencies, the spectrum they observe does not appears to 
come from overtones of a \B--\X{} band.
The only experimental reference to a $^1\Pi$ state in this energy range is from 
\cite{88SiMiHeHa.ZrO} who explain perturbation in the \Te{} triplet splittings 
using a $^1\Pi$ state originally predicted theoretically by 
\cite{69Green.ZrO}. The energies of electronic states in this energy range for 
transition metal diatomics are notoriously challenging to predict accurately even 
with today's methods \citep{jt632} and thus this early theoretical investigation cannot be trusted even qualitatively for higher electronic states, especially since more recent theoretical papers 
\citep{90LaBa.ZrO} make no such predictions for a $^1\Pi$ state in this energy 
range.

Attempts to assign the clearly visible bands seen by \cite{10BaCh.ZrO} to a 
\ZrO{} transition were unconvincing. Given the low resolution of these data and 
its inconsistencies with current knowledge of the electronic structure of ZrO 
from both a theoretical and experimental perspective, we suggest these 
unassigned peaks are due to \ce{ZrO+}. The method used by \citet{10BaCh.ZrO}  
does involve the creation of ions, and there is precedence for this occurring.  
\citet{79PhDa.ZrO+} conducted a study on bands in what was understood to be the 
ZrO spectrum with heads at 7811 and 8192 \AA{} that had previously been observed by 
\cite{50Afaf.ZrO}  and analysed by \citet{55UlAA.ZrO} and \citet{56UhAk.ZrO} as belonging to 
a new system. They found these bands belonged to a $^2\Pi$ - $^2\Sigma$ system 
of \ce{ZrO+}. While further work has been done on \ce{ZrO+}, none of these 
studies have examined the same wavelengths as \cite{10BaCh.ZrO}, and thus no 
definitive assignment can be made at this stage.

We do not extensively review the theoretical literature, but 
notable calculations include those of \citet{88LaBa.ZrO},
\citet{90LaBa.ZrO} and \citet{11ShSr.ZrO}.

\subsection{\label{subsec:comments} Rotationally-resolved data sources}
Our analysis started by digitising available assigned rovibronic transitions data, then converting them to \Marvel\ format. The full list of 
compiled data converted to \Marvel\ format is given in the Supplementary Information; 
an extract is given in \Cref{tab:marvelinput}. The full list of data sources used in the rotationally-resolved \Marvel\ analysis are summarised in \Cref{tab:datasources}; we provide information on the vibronic bands measured, the wavenumber range and  $J$ range, as well as the number of transitions measured. In total, we  use 12 data sources, involving \noelec{} electronic states with \notrans{} transitions and \nospinvibronic{} total unique spin-vibronic bands (ignoring $\Lambda$ splitting). 






Comments related to \Cref{tab:datasources}, particularly regarding the initial uncertainty chosen for the data, are as follows:

\begin{description}
\item[\underline{54LaUhBa:}] An uncertainty of 0.1 \cm{} was chosen,  enabling 
a high number of validated transitions within this dataset and those by the same 
author in the same year. 
\item[\underline{54Uhler:}] Uncertainty as for 54LaUhBa, though this data set 
could be compared against later data more directly and thus had a bigger 
influence on setting the maximum uncertainty used. 
\item[\underline{56UhAk:}] Uncertainty as for 54LaUhBa. The symmetry of the two 
electronic states was not confirmed at the time of publication, but both can now be assigned 
as $^1\Sigma^+$, given the later assignment of the ground state symmetry as 
$^1\Sigma^+$ and the lack of a Q branch. 
\item[\underline{57Akerlind:}] Uncertainty as for 54LaUhBa; this data were the 
only source of \F{} state information, so uncertainty reflects only requirements 
for self-consistency within this data set. 
\item[\underline{73BaTa:}] The data table has poor readability and it is likely 
that minor errors in digitisation may exist, though major errors were corrected 
by hand using the systematic nature of the transition frequencies.  We adopted 
0.01 \cm{} as the minimum uncertainty for the data(with higher values adopted 
as necessary up to 0.16 \cm{}), as this yielded a reasonable number of 
self-consistent results.
\item[\underline{73Lindgren:}] No uncertainty is stated in the paper; however, 
0.07 \cm{} gave reasonable self-consistent calculations for most bands. Note 
that these are satellite bands and hence had lower intensities and higher position uncertainties than for the main bands.  
\item[\underline{76PhDa.BX, 76PhDa.CX:}] The original paper did not use the \C{} label for 
the upper state; this has been named in subsequent discussions of \ZrO\ and 
adopted here. There are no uncertainties given; however, we found that at least 
0.02 \cm{} was required to enable a significant number of this data to self validate. 
 Some data were substantially more inaccurate than this; we have removed 
all data that required uncertainties of more than 0.2 \cm{} to be consistent 
with the rest of the data. 
\item[\underline{79PhDa:}] Data obtained from Kurucz and given uncertainties of 
0.02 \cm{}, as for other Phillips and Davis data of this era. The \Td$_{3}$ -- 
\Ta$_2$(0-1) band appears to be largely incorrectly assigned; we have used only 
those transitions that agree well with assignments from other bands.
\item[\underline{80HaDa, 81HaDaZo, 81HaDa:}] 0.01 \cm{} was stated as the measurement accuracy for 
at least some bands; this was adopted for the whole data set by multiple papers by the same authors in similar time period. Note that this is a factor 
of two more accurate than earlier data from Davis and co-workers. 
\item[\underline{88SiMiHuHa:}] The stated uncertainty was 200 MHz, with 
reproducability to 50 MHz; we therefore adopted 0.006 \cm{} as an initial 
uncertainty for our data. 
\item[\underline{90SuLoFrMa:}] The stated uncertainty was 4 kHz 
(3$\times$10$^{-8}$ \cm{}); however, consistency with 99BeGe 
required an uncertainty estimate of 3$\times$10$^{-7}$ \cm{}. 
\item[\underline{95Jonsson:}] The stated uncertainty was 0.016 \cm{}; however, 
we found a smaller uncertainty of 0.006 \cm{} as an initial estimate was 
warranted due to the consistency of the data both internally and with other 
results. 
\item[\underline{95KaMcHe:}] The stated uncertainty was 0.03 \cm{}; however, we 
found a smaller uncertainty of 0.007 \cm{} as an initial estimate was warranted 
due to the consistency of the data both internally and with other results. 
\item[\underline{99BeGe:}] The stated uncertainty was 1 kHz(10$^{-8}$ \cm{}); 
however, consistency with 90SuLoFrMa required uncertainty of 
10$^{-7}$ \cm{}, so this was adopted for all values.  
\end{description}

During the \Marvel{} process, many of our initial estimated
uncertainties were updated to establish a self-consistent network,
while some transitions were removed from consideration (designated
through a minus sign at the start of the \Marvel{} input line for that
transition). To assess the data, \Cref{tab:datasources} provides data
on the minimum, average and maximum uncertainty of each transition; in
most cases, we were able to keep the minimum and average uncertainty
to within a factor of two. We validated \novalid{} of our \notrans{}
input transitions, i.e. showed that these validated transitions were
consistent with other measurements. The \ZrO{} \Marvel{} input file
can readily be updated in the future with new spectroscopic
information to enable an updated set of \Marvel{} energies to be
created.



  \begin{table}
\caption{\label{tab:results} Extract from the 90Zr-16O.main.energies output file for \ZrO. Energies and their uncertainties are given in \cm{}. No indicates the number of transitions which contributed to the stated energy and uncertainty. }
\begin{center}
\renewcommand{\arraystretch}{1.2}
\begin{tabular}{lccllr}
\toprule
   \mc{1}{c}{State}  &  $v$  &    $J$    &    \mc{1}{c}{$\tilde{E}$}   &   \mc{1}{c}{Unc.}  &  \mc{1}{c}{No} \\
\midrule

X1Sigma+  & 5  & 92  & 	8286.730593	 & 0.016290	 & 3\\
a3Delta\_1  & 4  & 93  & 	8289.132156	 & 0.013142	 & 3\\
a3Delta\_2  & 4  & 89  & 	8295.024932	 & 0.018098	 & 3\\
X1Sigma+  & 6  & 79  & 	8296.993918	 & 0.009580	 & 6\\
a3Delta\_1  & 1  & 124  & 	8312.622601 & 	0.022336 & 	7\\
A1Delta  & 0  & 76  & 	8313.129649	 & 0.004368	 & 11\\
a3Delta\_2  & 5  & 76 	 & 8320.754165	 & 0.013995	 & 5\\
a3Delta\_2  & 3  & 101  & 	8322.052377	 & 0.013804 & 	6\\
a3Delta\_2  & 1  & 121  & 	8322.707720	 & 0.012197 & 	7\\
X1Sigma+  & 4  & 104  & 	8327.541550	 & 0.020000	 & 1\\
A1Delta &  1  & 60  & 	8336.081220 & 	0.008485	 & 2\\
a3Delta\_2 &  0  & 130  & 	8336.647124	 & 0.013307 & 	6 \\
\bottomrule
\end{tabular}
  \end{center}
  
\end{table}

\begin{center}
\footnotesize 
\begin{longtable}{lllclllllll}

\caption{\label{tab:SN}\textsc{Spectroscopic Network containing \noenergy{} energy levels 
 and their characteristics for \ZrO}}\\
 \toprule
 & $v$   &  $J$ Range   & \mc{3}{c}{Uncertainties (\cm{})} \\
\cline{4-6}
&  &  & Min & Av & Max \\
\midrule
\endfirsthead

\multicolumn{5}{c}%
{{ \tablename\ \thetable{} -- continued from previous page}} \\
\toprule
 & $v$   &  $J$ Range   & \mc{3}{c}{Uncertainties (\cm{})} \\
 \cline{4-6}
&  &  & Min & Av & Max \\
\midrule
\endhead

\bottomrule
\multicolumn{5}{c}{{Continued on next page}} \\ 
\endfoot

\bottomrule
\endlastfoot
\X$_{}$  &  0 & 0 - 131  & 1$\times 10^{-7}$ & 0.0096 & 0.055\\
 &  1 & 2 - 107  & 0.006 & 0.016 & 0.08\\
 &  2 & 2 - 115  & 0.0079 & 0.016 & 0.11\\
 &  3 & 2 - 105  & 0.008 & 0.014 & 0.052\\
 &  4 & 2 - 106  & 0.012 & 0.02 & 0.068\\
 &  5 & 2 - 107  & 0.0089 & 0.015 & 0.051\\
 &  6 & 2 - 107  & 0.0082 & 0.014 & 0.067\\
 &  7 & 3 - 66  & 0.012 & 0.016 & 0.042\\
\vspace{-0.5em}\\ 
\A$_{}$  &  0 & 2 - 133  & 0.0031 & 0.0058 & 0.12\\
 &  1 & 3 - 115  & 0.0045 & 0.012 & 0.14\\
\vspace{-0.5em}\\ 
\B$_{e}$  &  0 & 1 - 133  & 0.0043 & 0.0075 & 0.065\\
 &  1 & 1 - 116  & 0.0038 & 0.011 & 0.057\\
 &  2 & 1 - 117  & 0.0048 & 0.011 & 0.053\\
 &  3 & 1 - 108  & 0.0084 & 0.015 & 0.062\\
 &  4 & 1 - 116  & 0.0097 & 0.02 & 0.21\\
 &  5 & 2 - 67  & 0.01 & 0.018 & 0.045\\
\vspace{-0.5em}\\ 
\B$_{f}$  &  0 & 1 - 133  & 0.0046 & 0.0089 & 0.15\\
 &  1 & 1 - 118  & 0.0053 & 0.012 & 0.052\\
 &  2 & 2 - 115  & 0.0055 & 0.013 & 0.076\\
 &  3 & 2 - 106  & 0.012 & 0.02 & 0.08\\
 &  4 & 2 - 107  & 0.012 & 0.023 & 0.14\\
 &  5 & 3 - 66  & 0.014 & 0.019 & 0.042\\
\vspace{-0.5em}\\ 
\C$_{}$  &  0 & 0 - 121  & 0.0041 & 0.02 & 0.21\\
\vspace{-0.5em}\\ 
\F$_{}$  &  0 & 17 - 102  & 0.057 & 0.063 & 0.14\\
 &  1 & 35 - 93  & 0.057 & 0.064 & 0.1\\
\vspace{-0.5em}\\ 
\Ta$_{1}$  &  0 & 2 - 150  & 0.0026 & 0.0072 & 0.063\\
 &  1 & 2 - 150  & 0.007 & 0.014 & 0.12\\
 &  2 & 1 - 150  & 0.0074 & 0.013 & 0.084\\
 &  3 & 2 - 150  & 0.0072 & 0.013 & 0.059\\
 &  4 & 3 - 150  & 0.012 & 0.021 & 0.12\\
 &  5 & 2 - 106  & 0.0089 & 0.017 & 0.06\\
\vspace{-0.5em}\\ 
\Ta$_{2}$  &  0 & 2 - 149  & 0.0023 & 0.0086 & 0.057\\
 &  1 & 1 - 148  & 0.0052 & 0.014 & 0.08\\
 &  2 & 1 - 150  & 0.0072 & 0.017 & 0.2\\
 &  3 & 2 - 147  & 0.0075 & 0.018 & 0.17\\
 &  4 & 3 - 150  & 0.012 & 0.021 & 0.15\\
 &  5 & 2 - 136  & 0.0089 & 0.016 & 0.054\\
\vspace{-0.5em}\\ 
\Ta$_{3}$  &  0 & 3 - 144  & 0.0024 & 0.013 & 0.2\\
 &  1 & 3 - 144  & 0.0069 & 0.015 & 0.15\\
 &  2 & 3 - 145  & 0.0071 & 0.012 & 0.04\\
 &  3 & 3 - 135  & 0.0077 & 0.012 & 0.062\\
 &  4 & 3 - 105  & 0.012 & 0.017 & 0.043\\
 &  5 & 3 - 130  & 0.0082 & 0.014 & 0.055\\
\vspace{-0.5em}\\ 
\Tb$_{0e}$  &  0 & 7 - 106  & 0.0035 & 0.0065 & 0.023\\
\vspace{-0.5em}\\ 
\Tb$_{0f}$  &  0 & 14 - 91  & 0.0042 & 0.0064 & 0.037\\
\vspace{-0.5em}\\ 
\Tb$_{1e}$  &  0 & 11 - 100  & 0.0035 & 0.0056 & 0.013\\
\vspace{-0.5em}\\ 
\Tb$_{1f}$  &  0 & 20 - 90  & 0.0036 & 0.0058 & 0.012\\
\vspace{-0.5em}\\ 
\Tb$_{2e}$  &  0 & 12 - 111  & 0.0035 & 0.0051 & 0.015\\
\vspace{-0.5em}\\ 
\Tb$_{2f}$  &  0 & 12 - 111  & 0.0035 & 0.0053 & 0.031\\
\vspace{-0.5em}\\ 
\Td$_{2}$  &  0 & 2 - 150  & 0.0081 & 0.014 & 0.14\\
 &  1 & 2 - 151  & 0.0071 & 0.013 & 0.22\\
 &  2 & 2 - 151  & 0.0075 & 0.013 & 0.056\\
 &  3 & 3 - 151  & 0.0069 & 0.011 & 0.035\\
 &  4 & 3 - 106  & 0.0086 & 0.014 & 0.041\\
\vspace{-0.5em}\\ 
\Td$_{3}$  &  0 & 3 - 148  & 0.01 & 0.018 & 0.12\\
 &  1 & 2 - 151  & 0.0067 & 0.011 & 0.046\\
 &  2 & 2 - 147  & 0.0071 & 0.015 & 0.13\\
 &  3 & 2 - 150  & 0.0063 & 0.012 & 0.09\\
 &  4 & 3 - 136  & 0.0088 & 0.014 & 0.029\\
\vspace{-0.5em}\\ 
\Td$_{4}$  &  0 & 4 - 144  & 0.0081 & 0.016 & 0.11\\
 &  1 & 4 - 145  & 0.0072 & 0.013 & 0.053\\
 &  2 & 4 - 145  & 0.0072 & 0.014 & 0.16\\
 &  3 & 4 - 145  & 0.0068 & 0.013 & 0.12\\
 &  4 & 4 - 131  & 0.0082 & 0.012 & 0.037\\
\vspace{-0.5em}\\ 
\Te$_{0e}$  &  0 & 15 - 59  & 0.058 & 0.07 & 0.1\\
\vspace{-0.5em}\\ 
\Te$_{0f}$  &  0 & 27 - 59  & 0.058 & 0.064 & 0.15\\
\vspace{-0.5em}\\ 
\Te$_{1e}$  &  0 & 3 - 73  & 0.0043 & 0.02 & 0.1\\
\vspace{-0.5em}\\ 
\Te$_{1f}$  &  0 & 3 - 67  & 0.0044 & 0.016 & 0.13\\
\vspace{-0.5em}\\ 
\Te$_{2}$  &  0 & 2 - 85  & 0.0043 & 0.034 & 0.1\\
\vspace{-0.5em}\\ 
\Tf$_{1}$  &  0 & 20 - 76  & 0.071 & 0.074 & 0.1\\
\vspace{-0.5em}\\ 
\Tf$_{2}$  &  0 & 20 - 79  & 0.071 & 0.075 & 0.1\\
\vspace{-0.5em}\\ 
\Tf$_{3}$  &  0 & 21 - 81  & 0.071 & 0.077 & 0.1\\

\vspace{-0.5em}\\ 

\end{longtable}
\end{center}

\begin{figure*}%
\includegraphics[width=0.9\textwidth]{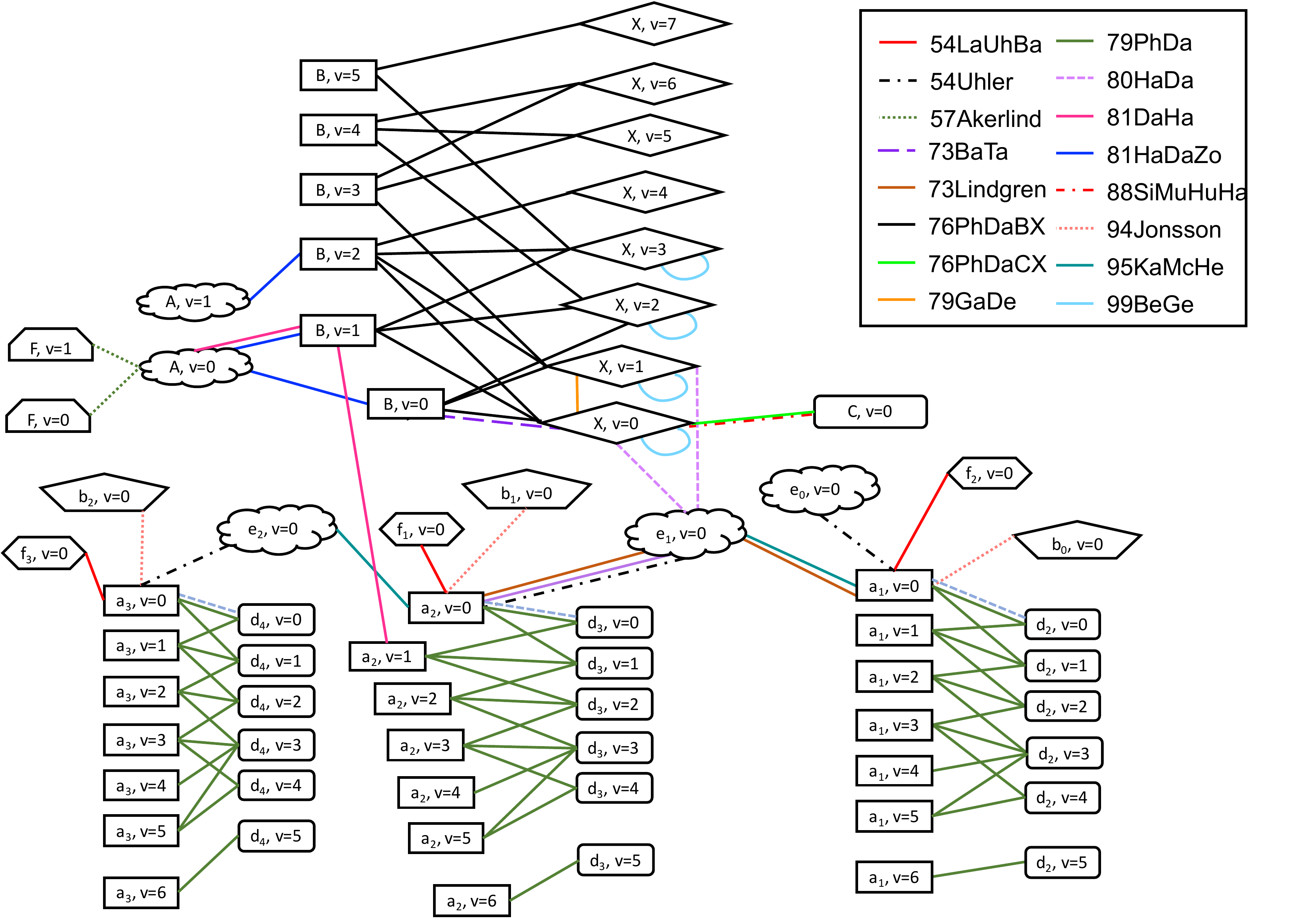}
\caption{\label{fig:SN}Depiction of connectivity of experimentally observed \ZrO\ bands.}
\end{figure*}

\begin{figure}
\includegraphics[width=0.5\textwidth]{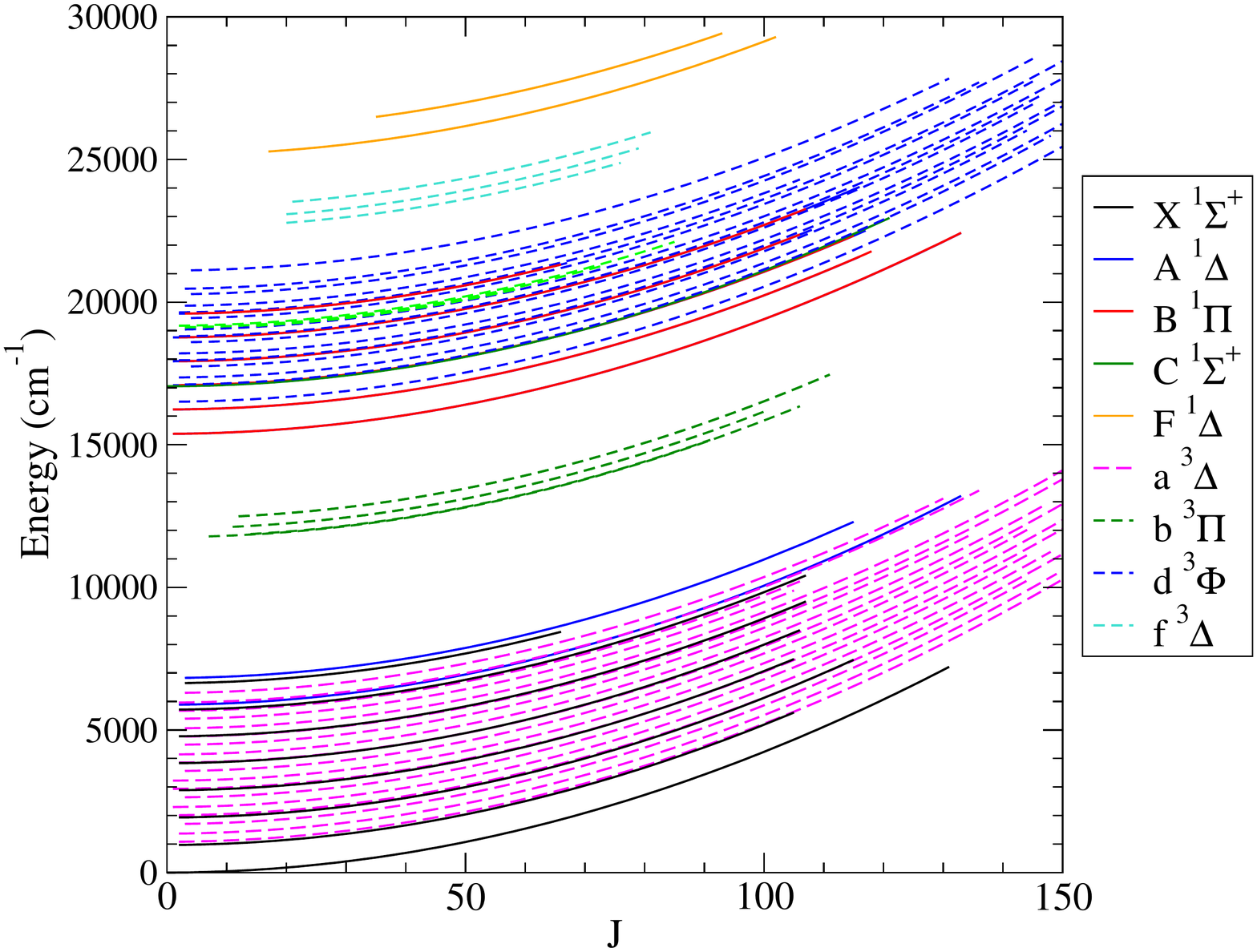}
\caption{\label{fig:ZrOenergylevels}\ZrO\ energy levels from \Marvel{} analysis.}
\end{figure}

\section{Results and Discussion}

\subsection{Spectroscopic Networks}
\Cref{fig:SN} represents the data from \Cref{tab:datasources} 
 showing the experimentally-measured transitions connecting
different vibronic states. From this diagram, it is clear that
there are three bands connecting the singlet and triplet manifold, and
some satellite transitions for the triplet sub-manifolds,
allowing most energy levels to be connected.


There are three minor free-floating networks connecting the \Ta($\nu=6$) and \Td($\nu=5$) levels. These could be connected through observing additional vibrational transitions, but this is not essential for producing a good model of the \ZrO{} electronic states.  

\subsection{Energy Levels}

\Cref{tab:results} shows an extract of the final empirical energy levels produced by \Marvel{} for \ZrO{} in this work. This list of energy levels includes an estimate for the uncertainty in the provided energy of the quantum state, as well as identifying the number of transitions used to determine the energy level; on average 5.3 transitions were used to find each energy level. 

\Cref{tab:SN} summarises the \noenergy{} empirical energy levels found in the main spectroscopic network from the \Marvel{} analysis for \ZrO. We see the minimum, average and maximum uncertatinty provided for the empirical energies from the \Marvel{} analysis. The minimum is usually very small, often less than 0.01 \cm{}, while the maximum can exceed 0.1 \cm{}; this is probably for higher  $J$  states. 
There is generally coverage to high rotational number  $J$  if the spin-vibronic level is known. 

These results show that there is good rotationally-resolved empirical
understanding of a reasonable number of vibrational states of the
\X{}, \B{}, \Ta{} and \Tb{} electronic states (sufficient for a good
potential energy curve to be fitted). However, there is much less
vibrational information (only one or two levels) for the \A{}, \C{},
\F{}, \Tb{}, \Te{} and \Tf{} states. This will cause significant
problems when fitting potential energy curves for a full spectroscopic
model and eventual line list for \ZrO{} and its isotopologues,
particularly for the \C{}, \Tb{}, \Te{}, \Tf{} states in which only
one vibrational level is known. Note that line lists of all
isotopologues can be easily produced using variational nuclear-motion 
techniques using
data from only a single isotopologue with reasonably high accuracy,
however {\it ab initio} predictions of vibrational constants especially for higher lying
electronic states of transition-metal-containing diatomics still have
errors of up to 50 \cm{} \citep{jt632}.

\Cref{fig:ZrOenergylevels} shows the empirical energy levels for the main spectroscopic network from \Marvel{} against  $J$  for each spin vibronic band. These are clearly quadratic and smooth, indicating there are no major problems with the \Marvel{} network for \ZrO.

\subsection{Comparison with \citet{PlezZrO}}

\begin{figure*}
\includegraphics[width=0.5\textwidth]{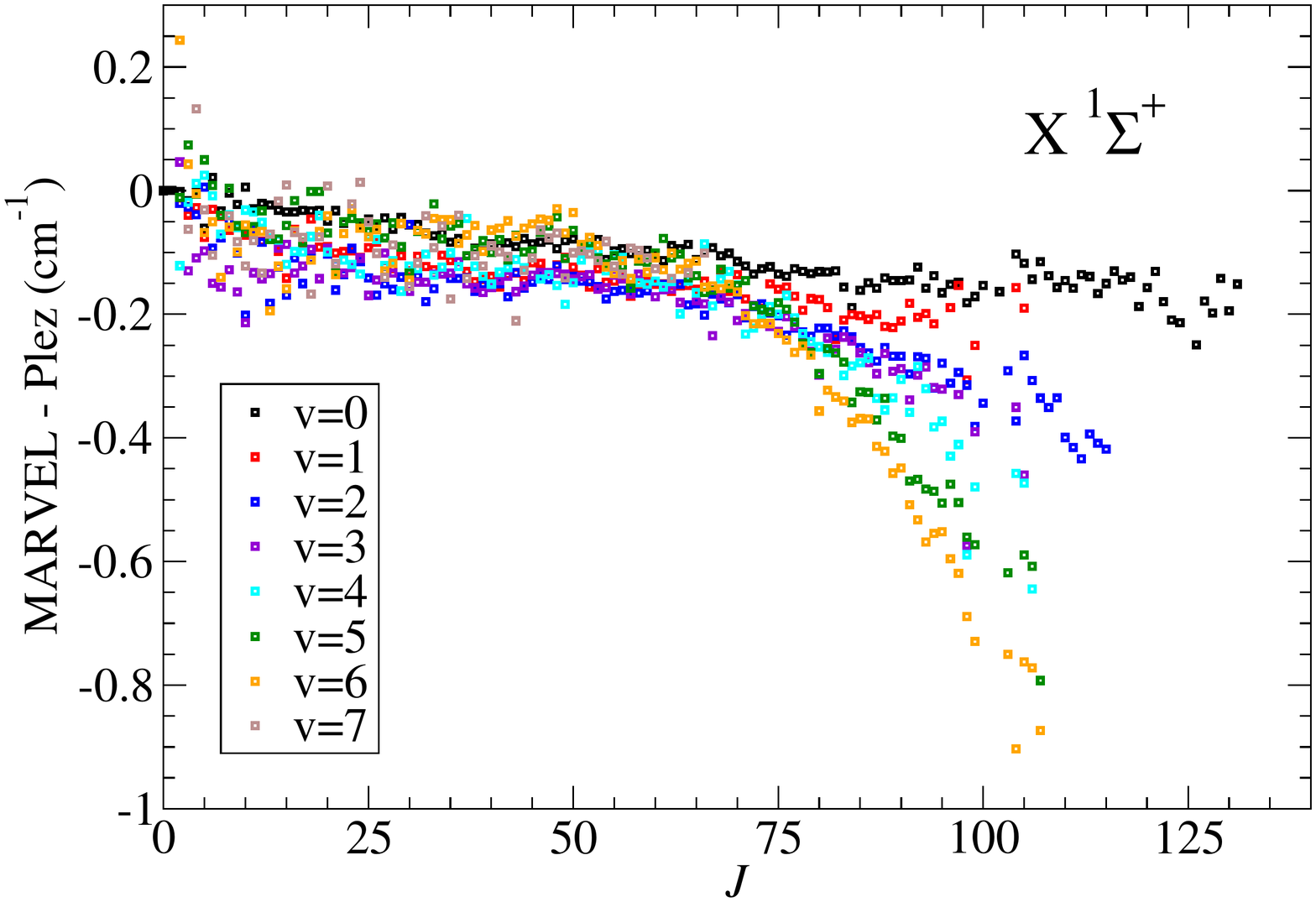}
\includegraphics[width=0.5\textwidth]{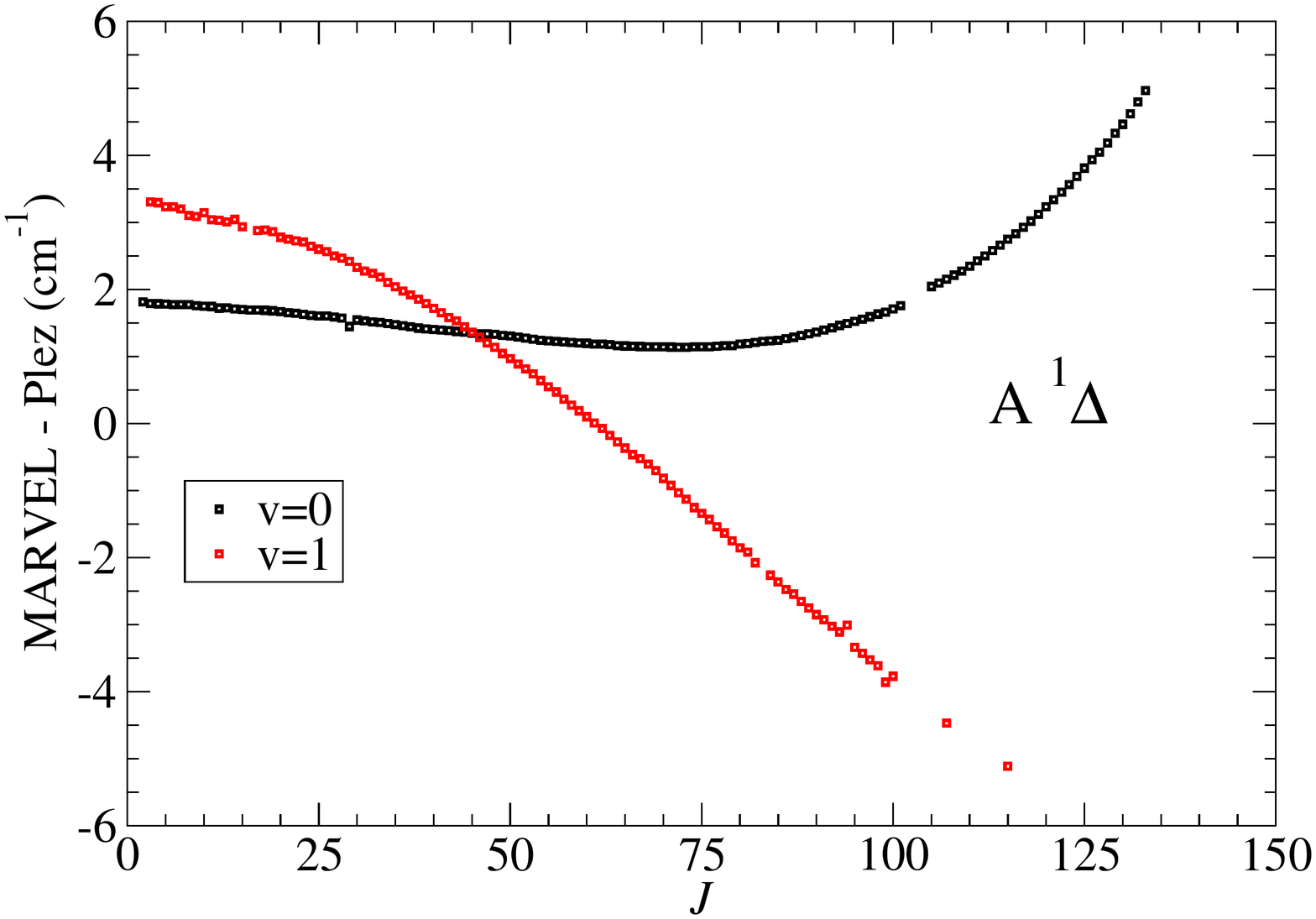}

\includegraphics[width=0.5\textwidth]{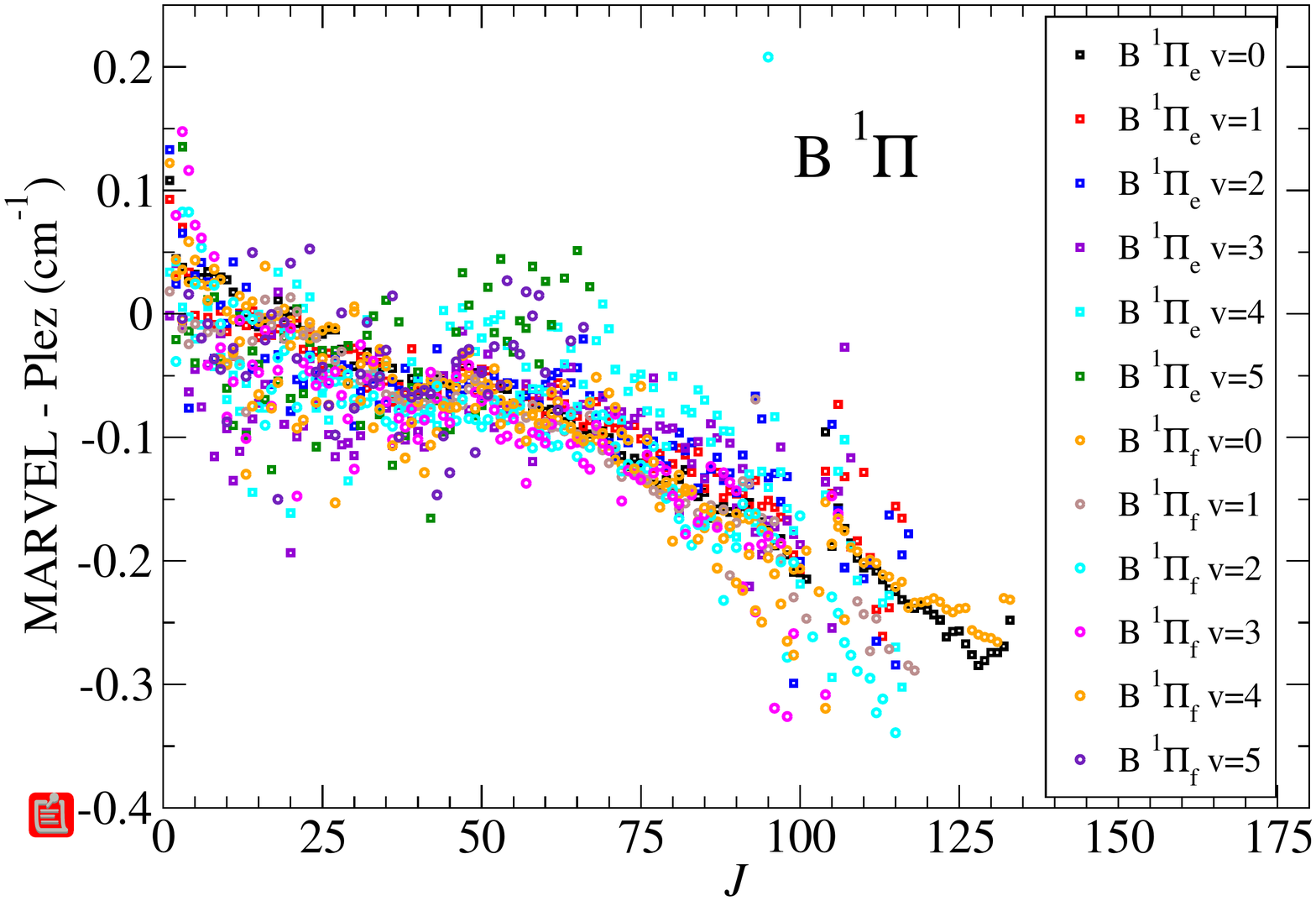}
\includegraphics[width=0.5\textwidth]{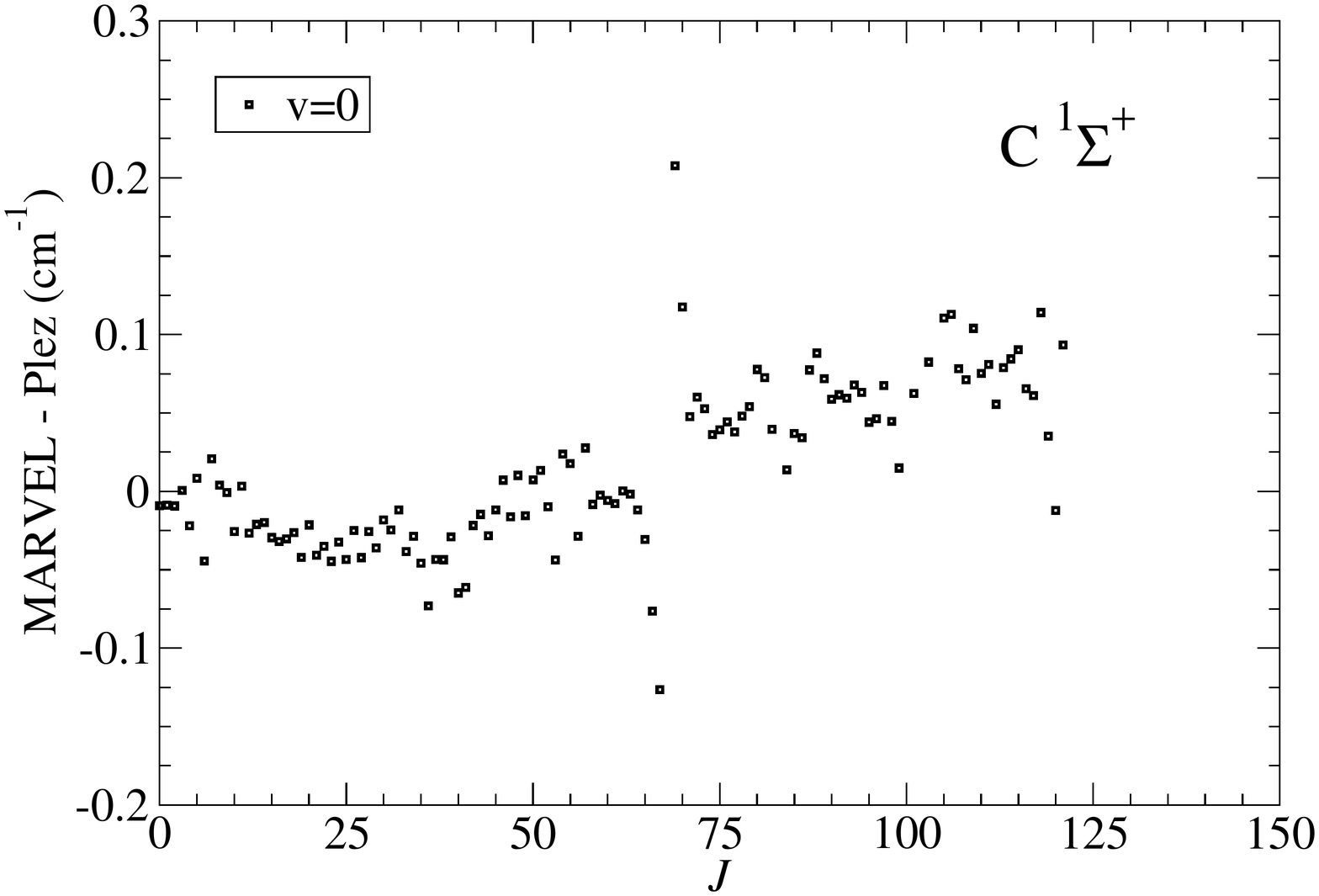}
\caption{\label{fig:Singleterror} Difference between \Marvel{} and \cite{PlezZrO} energy levels for the singlet states.}
\end{figure*}

\Cref{fig:Singleterror} shows the difference between the singlet \Marvel{} energy levels for \ZrO{} and those from the \cite{PlezZrO} \ZrO\ linelist. For the \X{}, \B{} and \C{} states, the differences average around 0.05-0.15 \cm{}, with somewhat higher deviations up to 1 \cm{} for large  $J$  especially in the \C{} state. The scatter here is probably largely a reflection of inaccuracies in the MARVEL energy levels, though perturbations not considered in the Plez line list might also contribute. The \A{} state, however, shows much more significant deviations; the $v$=0 state is off by about 2 \cm{} up to  $J$ =100, with much more significant deviations for larger  $J$ . The $v$=1 state also shows substantial differences of up to 4 \cm{} that changes significantly with  $J$ .


\begin{figure}
\includegraphics[width=0.5\textwidth]{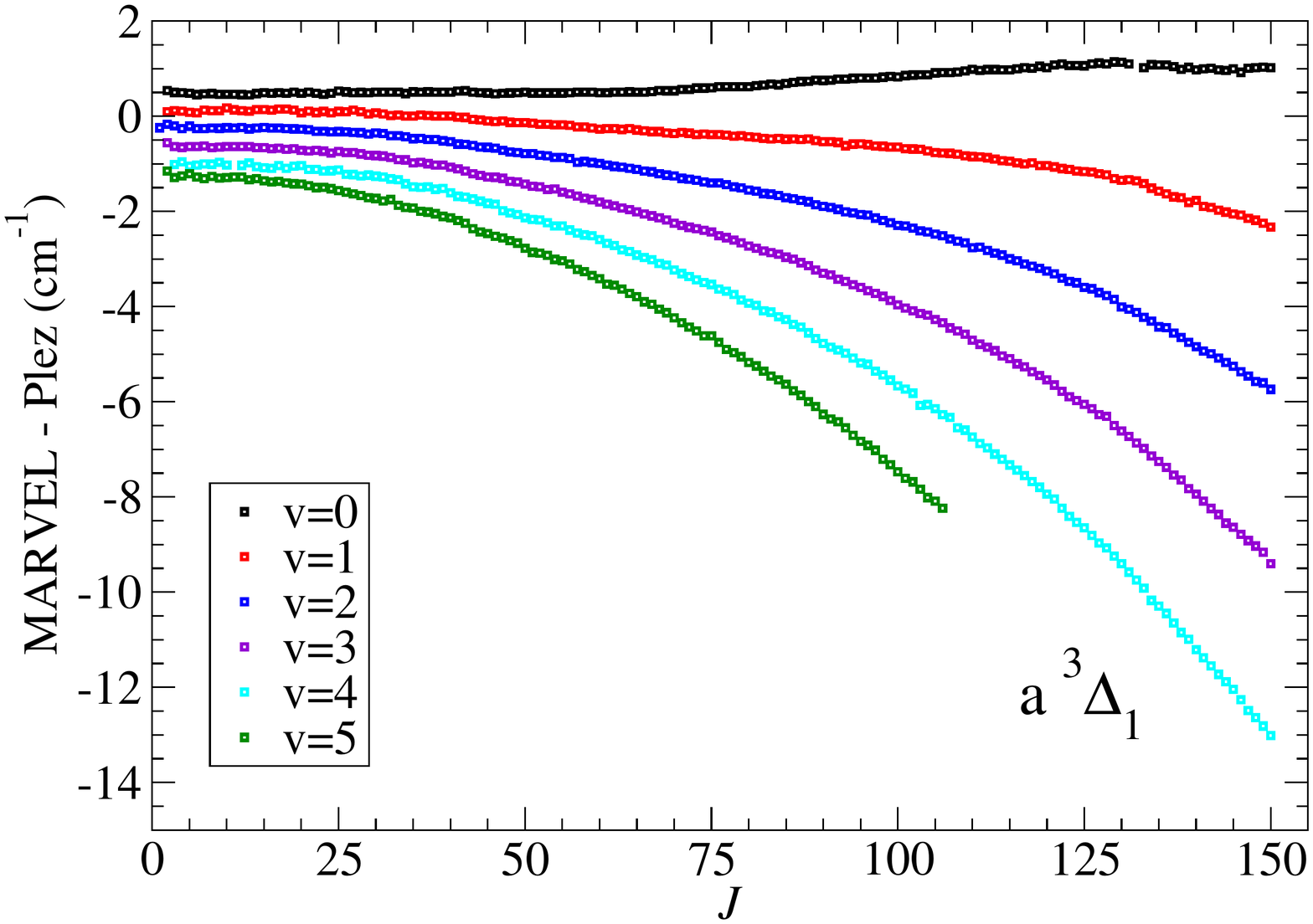}
\includegraphics[width=0.5\textwidth]{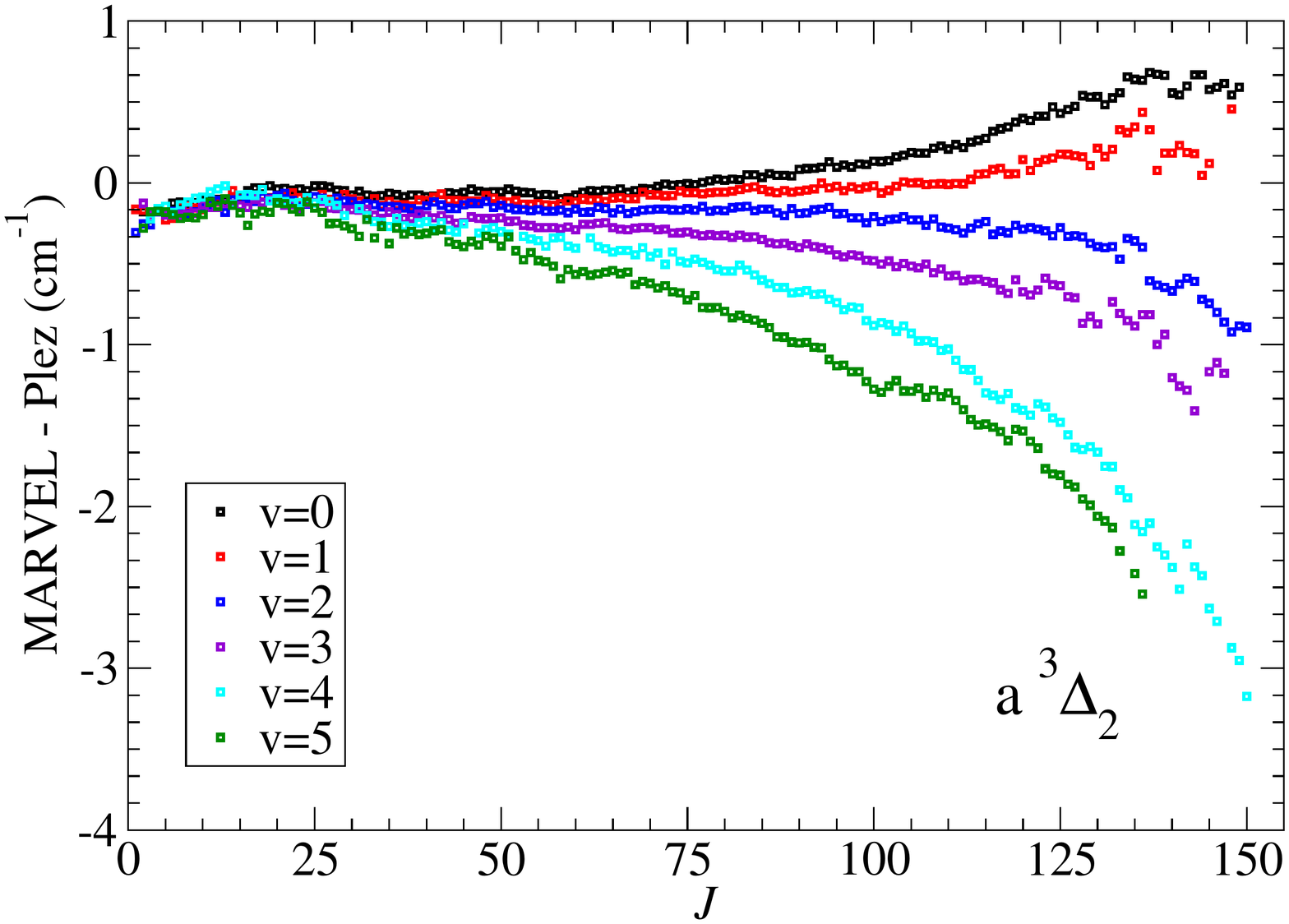}
\includegraphics[width=0.5\textwidth]{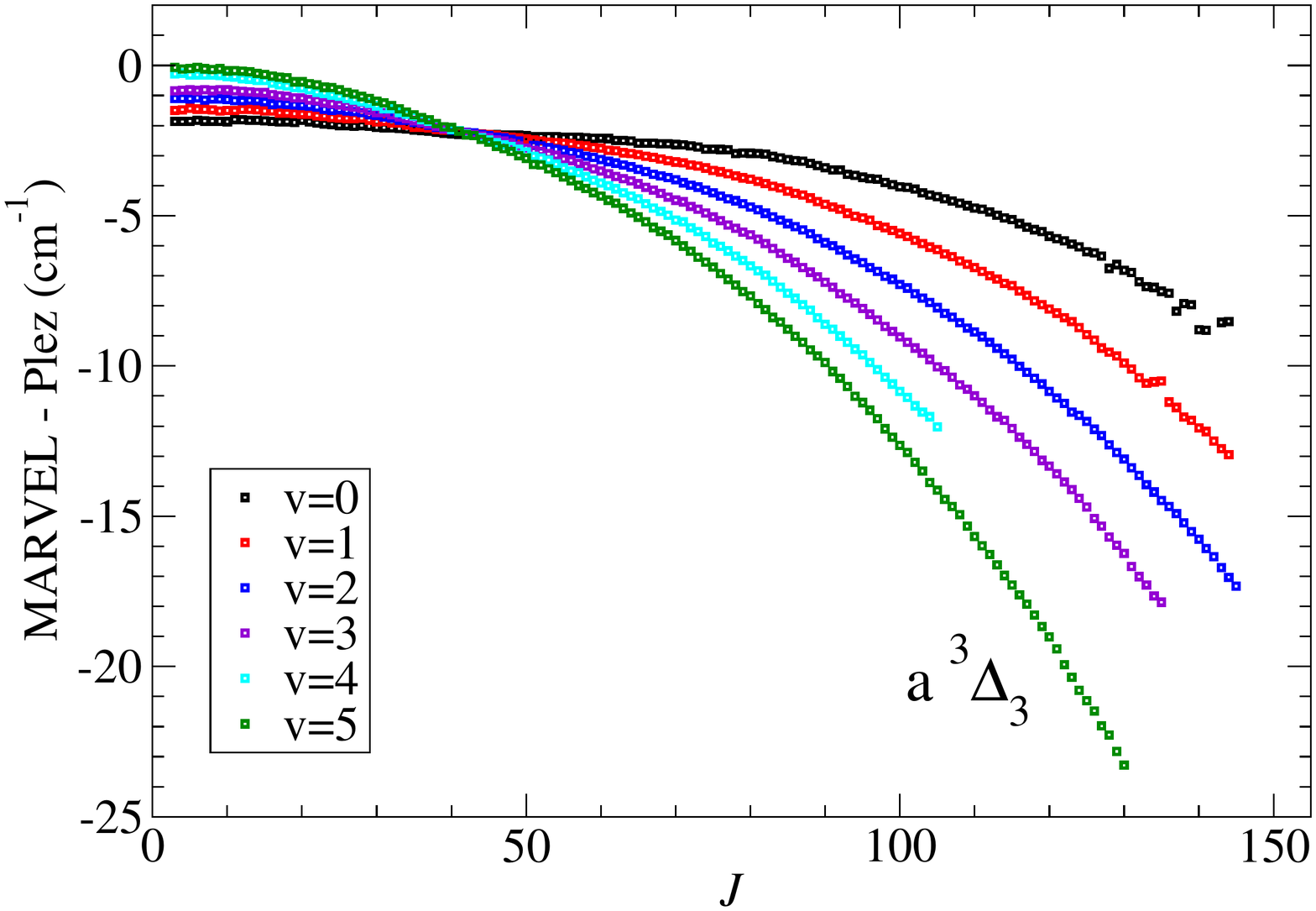}
\caption{\label{fig:a} Difference between \Marvel{} and \cite{PlezZrO} energy levels for the \Ta{} state.}
\end{figure}

\begin{figure}
\includegraphics[width=0.5\textwidth]{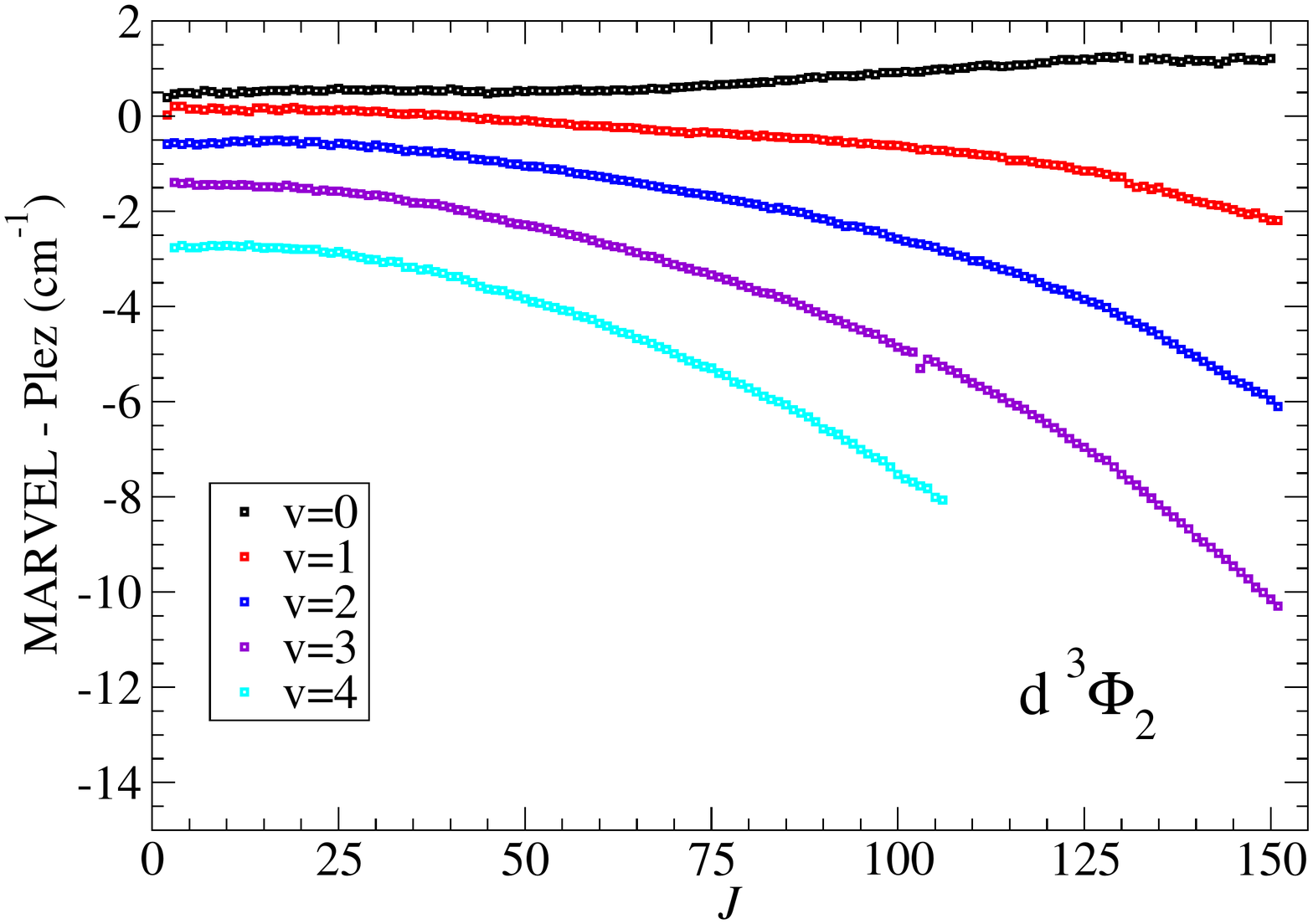}
\includegraphics[width=0.5\textwidth]{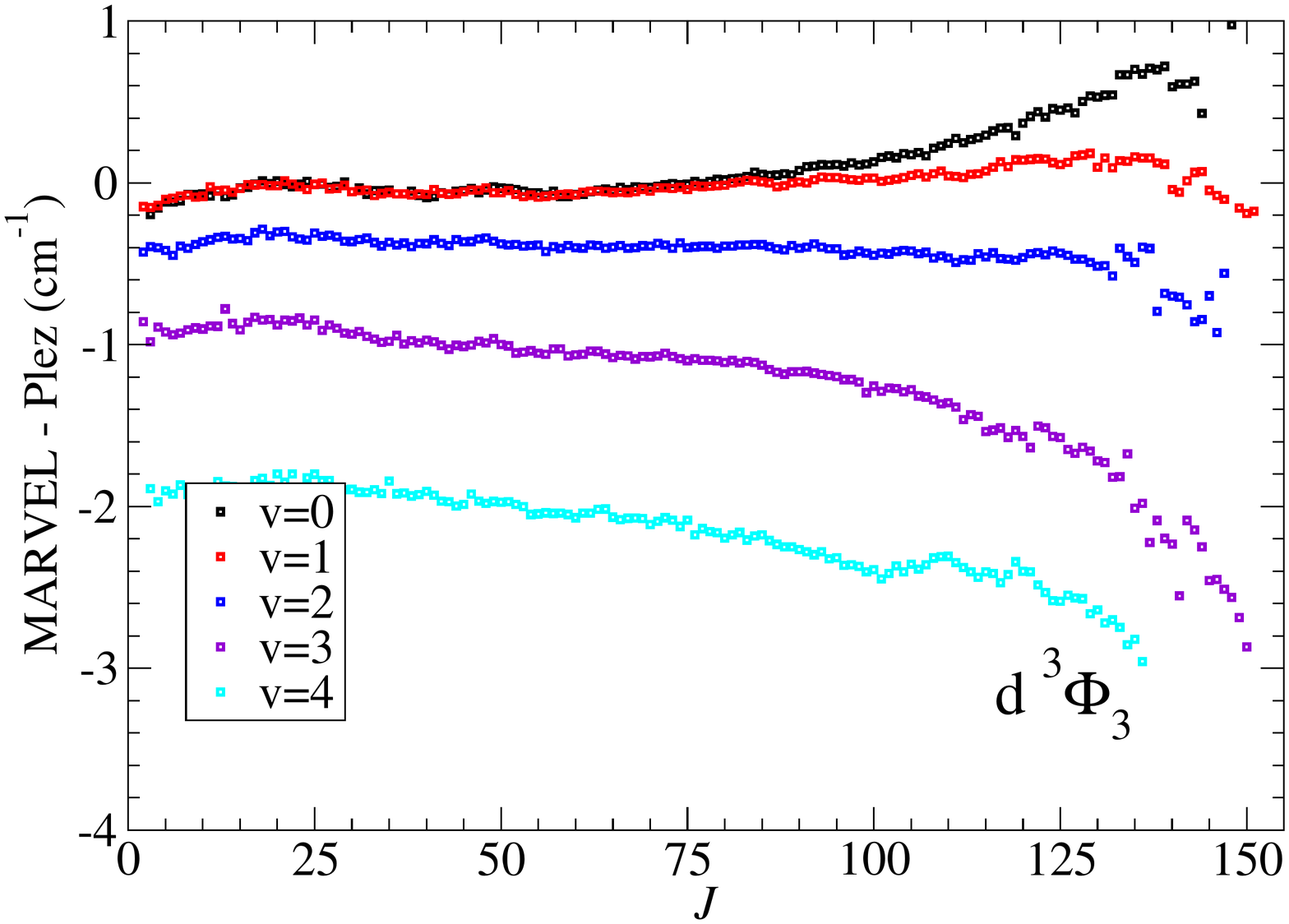}
\includegraphics[width=0.5\textwidth]{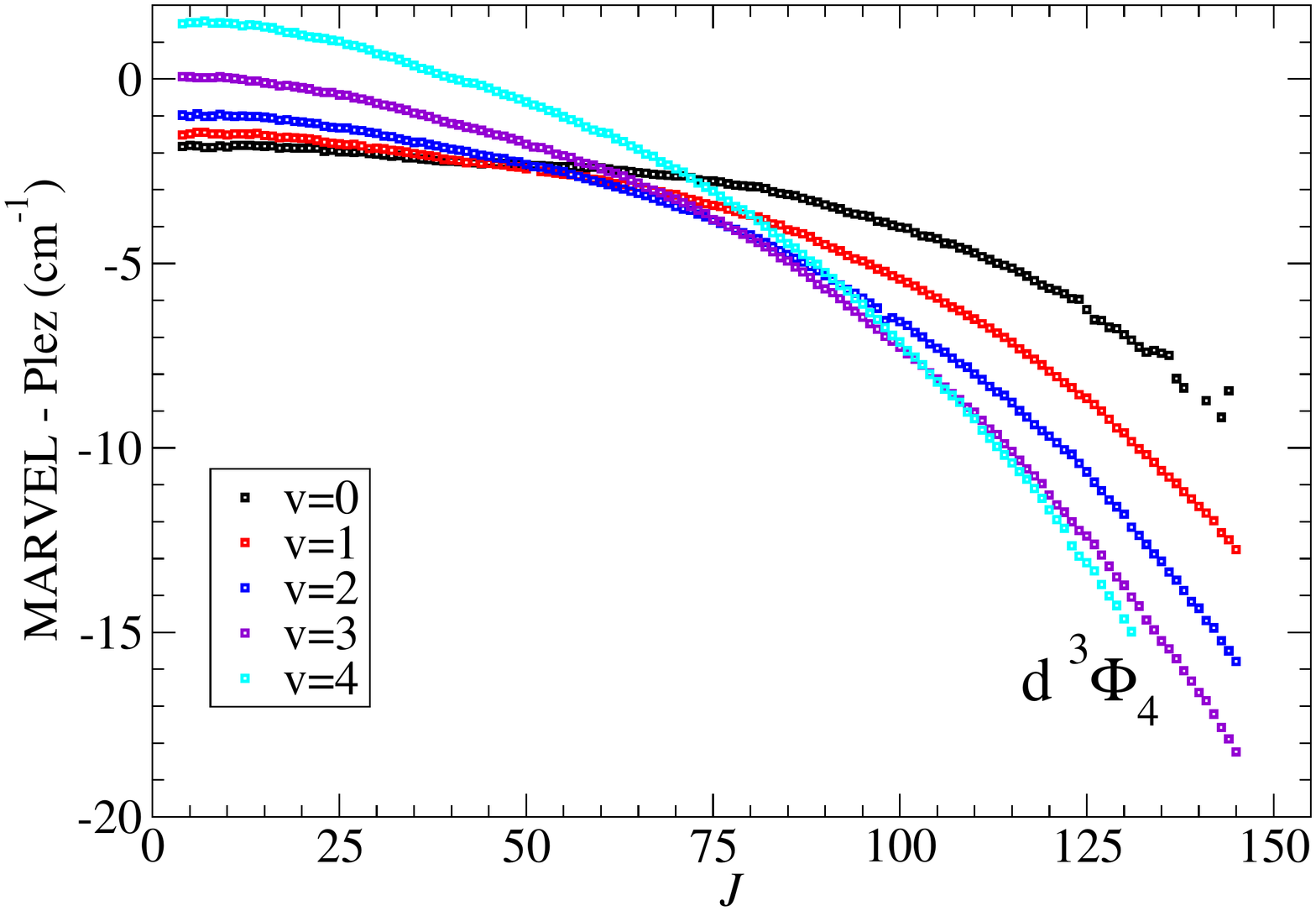}
\caption{\label{fig:d} Difference between \Marvel{} and \cite{PlezZrO} energy levels for the \Td{} state.}
\end{figure}

\begin{figure}
\includegraphics[width=0.5\textwidth]{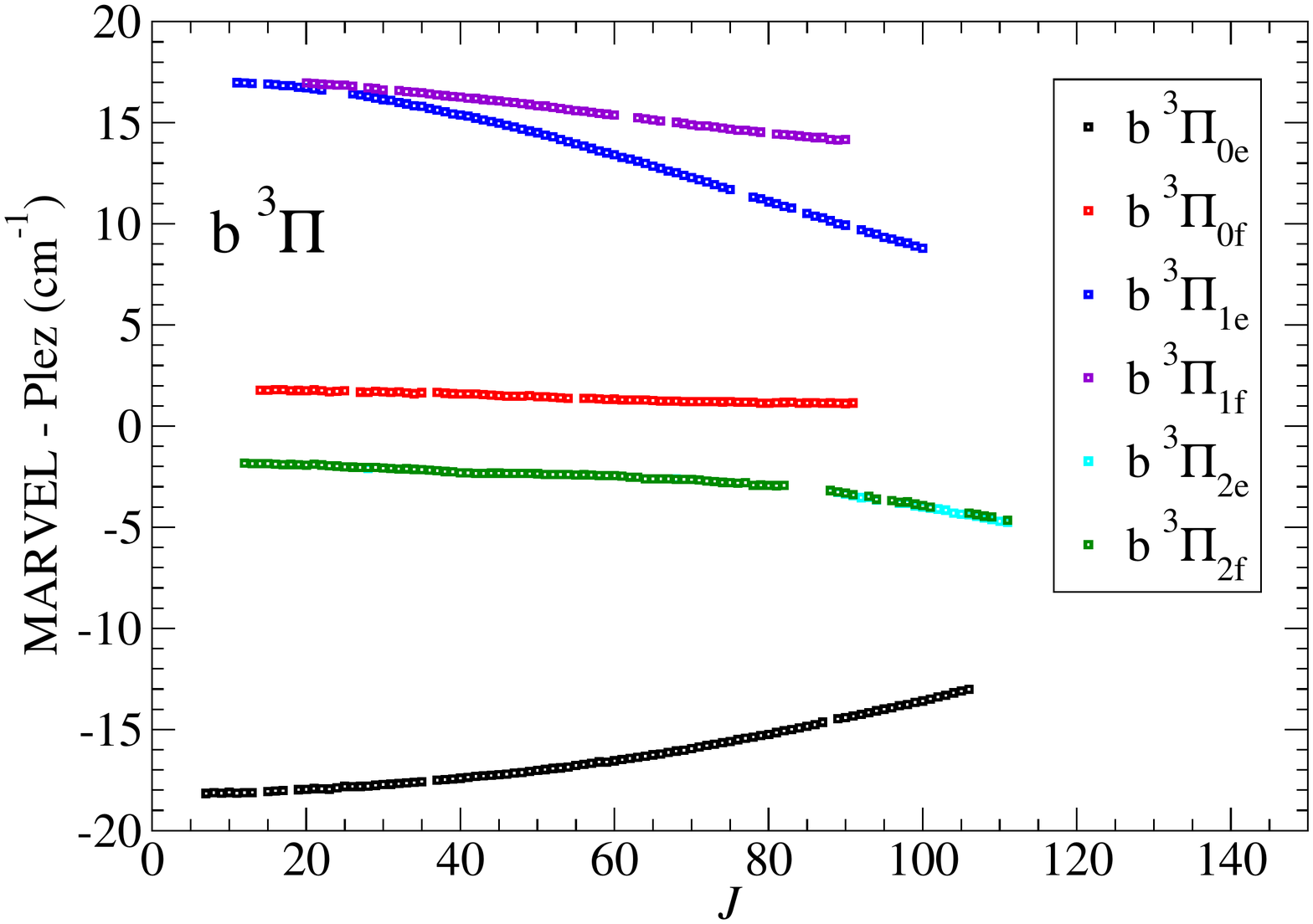}
\includegraphics[width=0.5\textwidth]{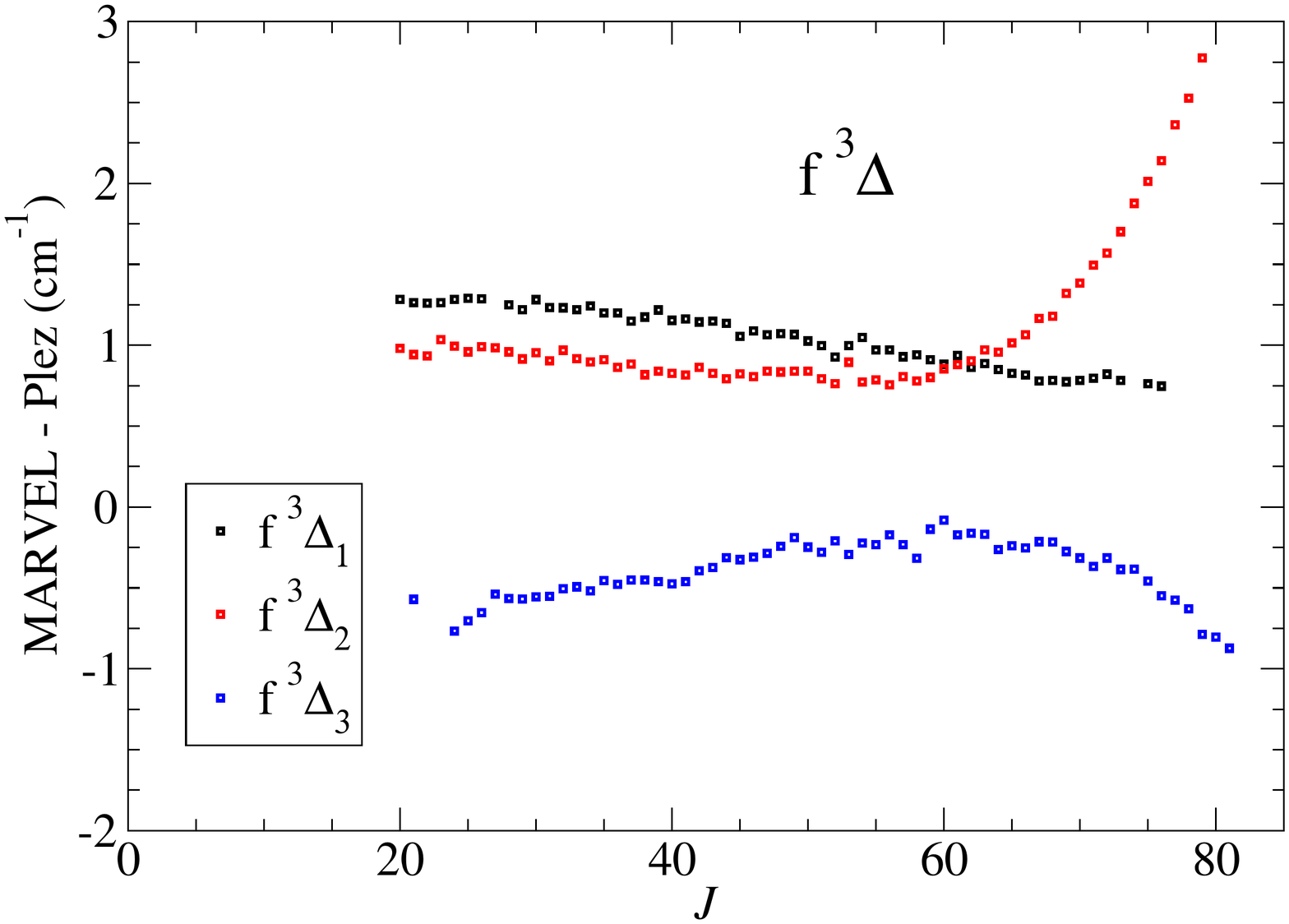}
\caption{\label{fig:f} Difference between \Marvel{} and \cite{PlezZrO} energy levels for the \Tb{} and \Tf{} state.}
\end{figure}

\Cref{fig:a,fig:d,fig:f} shows the difference between the triplet 
\Marvel{} energy levels for \ZrO{} and those from the \cite{PlezZrO} \ZrO\ 
linelist. These errors are much more significant than for the singlet states. 

Clear systematic errors can be seen throughout the \Ta{} and \Td{} bands - since 
these are a major cause of opacity of \ZrO{} in stellar conditions, improvements 
to these energy levels is highly desirable. Note, however, that since the errors 
in the \Ta$_{n}$ and \Td$_{n+1}$ parallel each other, the errors in transition 
frequencies in the Plez line list will be much smaller than the errors in 
energies that are plotted here. 

 The \Tb{} state shows significant and systematic errors in the Plez database compared to the \Marvel{} data of up to 15 \cm{} for many electronic states. Our adoption of the \citet{94Jonsson.ZrO} assignments in preference to the \cite{79PhDaGa.ZrO} assignments contributes to much larger lambda doubling in the \Marvel{} data than was adopted in the Plez data. There is also clear systematic differences in the energies of the \Tb$_{1}$ levels of more than 15 \cm{} in many regions. Smaller differences of up to 5 \cm{} were found in the \Tb$_2$ levels that parallel the errors in the \Ta$_3$ and \Td$_4$ energies, indicating that the errors associated with the transition frequencies in this band in the Plez line list will be much smaller.  

The \Tf{} data show that Plez's triplet 
splitting is in error by about 1 \cm{}, with some larger errors at 
high  $J$.

\begin{table}
    \caption{
    \label{tab:singletbandconstants}Spectroscopic band constants, in \cm, for the singlet vibronic bands, assuming no perturbations. }
    \centering
    \begin{tabular}{lcrrcH}
    \toprule
    \mc{1}{c}{$State$} &  \mc{1}{c}{$v$}  &  \mc{1}{c}{$T_v$}  &  \mc{1}{c}{$B_v$}  &  \mc{1}{c}{10$^7$ $D_v$} \\
    \midrule
\X		&	0	&	-0.028	&	0.4226	&	3.18	\\
	&	1	&	969.509	&	0.42065	&	3.19	\\
	&	2	&	1932.154	&	0.41868	&	3.18	\\
	&	3	&	2887.873	&	0.41674	&	3.22	\\
	&	4	&	3836.760	&	0.41475	&	3.22	\\
	&	5	&	4778.739	&	0.41275	&	3.21	\\
	&	6	&	5713.739	&	0.41077	&	3.23	\\
 \vspace{-0.5em} \\
\A	&	0	&	5887.160	&	0.41646	&	3.26	\\
&	1	&	6823.105	&	0.41457	&	3.26	\\
 \vspace{-0.5em} \\
\B 	&	0	&	15383.385	&	0.40151	&	3.51	\\
&	1	&	16236.949	&	0.39959	&	3.50	\\
&	2	&	17084.607	&	0.39765	&	3.51	\\
&	3	&	17926.299	&	0.39570	&	3.51	\\
&	4	&	18762.010	&	0.39375	&	3.52	\\
&	5	&	19591.668	&	0.39175	&	3.40	\\
 \vspace{-0.5em} \\
\C	&	0	&	17050.378	&	0.40480	&	3.44	\\
 \vspace{-0.5em} \\
\F	&	0	&	25159.631	&	0.39736	&	3.57	\\
&	1	&	25994.872	&	0.39540	&	3.66	\\
 \vspace{-0.5em} \\
\bottomrule
\end{tabular}
\end{table}

\begin{table*}
    \caption{
    \label{tab:tripletbandconstants}Spectroscopic constants, in \cm, for the triplet spin-vibronic band, assuming no perturbations. 
    }
    \centering
    \begin{tabular}{lcrrlrrlrrlrrrrr}
    \toprule 
 &   &  \mc{3}{c}{$\Sigma=-1$}  &  \mc{3}{c}{$\Sigma=0$}  &  \mc{3}{c}{$\Sigma=1$}  \\
\cmidrule(r){3-5}\cmidrule(r){6-8}\cmidrule(r){9-11} \cmidrule(r){12-13}

\mc{1}{c}{$v$}  &   &  \mc{1}{c}{$T_v$}  &  \mc{1}{c}{$B_v$}  &  \mc{1}{c}{$D_v$(10$^7$)} 
     &  \mc{1}{c}{$T_v$}  &  \mc{1}{c}{$B_v$}  &  \mc{1}{c}{$D_v$(10$^7$)} 
     &  \mc{1}{c}{$T_v$}   &  \mc{1}{c}{$B_v$}  &  \mc{1}{c}{$D_v$(10$^7$)}   &  \mc{1}{c}{$\Delta(SO)$}\\
\midrule
\cmidrule(r){3-5}\cmidrule(r){6-8}\cmidrule(r){9-11} \cmidrule(r){12-13}
 &   &  \mc{3}{c}{\Ta$_1$}  &  \mc{3}{c}{\Ta$_2$}  &   \mc{4}{c}{\Ta$_3$}  \\
\cmidrule(r){3-5}\cmidrule(r){6-8}\cmidrule(r){9-11} 
0	&	&	1080.363	&	0.41333	&	3.18	&	1367.750	&	0.41476	&	3.26				&	1703.505	&	0.41565	&	3.43	\\
1	&	&	2011.656	&	0.41144	&	3.19	&	2299.369	&	0.41285	&	3.26				&	2635.505	&	0.41374	&	3.44	\\
2	&	&	2936.474	&	0.40955	&	3.20	&	3224.476	&	0.41096	&	3.29				&	3561.029	&	0.41181	&	3.45	\\
3	&	&	3854.766	&	0.40765	&	3.21	&	4143.140	&	0.40903	&	3.28				&	4480.011	&	0.40989	&	3.45	\\
4	&	&	4766.602	&	0.40574	&	3.22	&	5055.317	&	0.40712	&	3.31				&	5392.650	&	0.40791	&	3.43	\\
5	&	&	5672.050	&	0.40378	&	3.19	&	5960.981	&	0.40516	&	3.28				&	6298.588	&	0.40602	&	3.48	\\
																					
\vspace{-0.5em} \\

 &  &  \mc{3}{c}{\Tb$_0$}  &  \mc{3}{c}{\Tb$_1$}  &   \mc{3}{c}{\Tb$_2$}  \\
\cmidrule(r){3-5}\cmidrule(r){6-8}\cmidrule(r){9-11} 
0	&	$e$		&	11765.173	&	0.40801	&	3.36	&	12069.846	&	0.40862	&	3.42	&	12427.697	&	0.40934	&	3.62	\\
&	$f$	&	11783.845	&	0.40754	&	3.29	&	12069.859	&	0.40915	&	3.45	&	12427.705	&	0.40933	&	3.60	\\

\vspace{-0.5em} \\
 &  &  \mc{3}{c}{\Td$_2$}  &  \mc{3}{c}{\Td$_3$}  &   \mc{3}{c}{\Td$_4$}  \\
\cmidrule(r){3-5}\cmidrule(r){6-8}\cmidrule(r){9-11} 
	0 &	&	16507.187	&	0.40312	&	3.57	&	17109.068	&	0.40368	&	3.60	&	17737.310	&	0.40430	&	3.77	\\
	1&	&	17357.358	&	0.40103	&	3.58	&	17958.303	&	0.40160	&	3.62	&	18588.627	&	0.40222	&	3.77	\\
	2&	&	18200.953	&	0.39894	&	3.59	&	18801.000	&	0.39949	&	3.63	&	19433.859	&	0.40015	&	3.78	\\
	3&	&	19038.014	&	0.39683	&	3.60	&	19637.131	&	0.39738	&	3.65	&	20273.306	&	0.39807	&	3.77	\\
	4&	&	19868.503	&	0.39469	&	3.58	&	20466.627	&	0.39524	&	3.62	&	21106.945	&	0.39602	&	3.77	\\
\vspace{-0.5em} \\
 &  &  \mc{3}{c}{\Te$_0$}  &  \mc{3}{c}{\Te$_1$}  &   \mc{3}{c}{\Te$_2$}  \\
\cmidrule(r){3-5}\cmidrule(r){6-8}\cmidrule(r){9-11} 
0	&	$e$	&	19074.117	&	0.39551	&	0.85	&	19113.069	&	0.40387	&	5.73			&	19169.508	&	0.40619	&	5.02	\\
&	$f$	&	19078.935	&	0.39449	&	0.04	&	19112.826	&	0.40266	&	4.79									\\			
	\vspace{-0.5em} \\
 &  &  \mc{3}{c}{\Tf$_1$}  &  \mc{3}{c}{\Tf$_2$}  &   \mc{3}{c}{\Tf$_3$}  \\
\cmidrule(r){3-5}\cmidrule(r){6-8}\cmidrule(r){9-11} 
0	& & 	22616.840	&	0.38945	&	3.43		&	22916.797	&	0.39207	&	0.67	&	23335.153	&	0.39572	&	3.01	\\
\bottomrule
    \end{tabular}
\end{table*}

\subsection{Band Constants}

Band constants were obtained by a quadratic fit of the energies of the lines against rotational quantum number $J$ for each band. 

\Cref{tab:singletbandconstants} shows the rotational band constants, $T_v$, 
$B_v$ and $D_v$ for each singlet vibronic bands. The centrifugal term, $D_v$ is 
reasonably constant within a given electronic state, while the rotational 
constants, $B_v$ decreases as expected as the bond length increases in higher 
vibrational states.

\Cref{tab:tripletbandconstants} shows the fitted rotational band constants for 
each spin-vibronic band in the triplet manifold for \ZrO{} from this \Marvel{} 
analysis.

A  compilation of band constants is given by
\cite{95KaMcHe.ZrO}; we find significant differences of 2 \cm{} in
the $T_0$ for \Td$_2$, \Td$_3$ and \Ta$_3$. We prefer our value,
however, as the \Td -- \Ta{} transitions form part of our input data,
whereas it is unclear how the \cite{95KaMcHe.ZrO} was compiled.
Otherwise, the $T_0$ values agree within 0.1 \cm{}.

\begin{table}
 
\centering
\caption{\label{tab:bh1}  R-branch  bandheads in \cm{} from the \X{} state for \ZrO;  $J$ gives the approximate $J$ value corresponding to the rotational transitions at the bandhead.}
\begin{tabular}{llrrlll}
\toprule
 &  $v$'-$v$"  &   $J$   &  MARVEL  &  Low-res obs. \\
\midrule
\B{} -- \X{}  &  0-0  &  18  &  15391.40 \\ 
   &  0-1  &  20  &  14422.64 \\
   &  0-2  &  22  &  13460.94 \\ 
   &  0-3*  &  25  &  12506.31 \\
   &  0-4*  &  29  &  11559.08 \\
   &  0-5*  &  34  &  10619.12 \\
   &  0-6*  &  41  &  9687.00 \\
   &  0-7*  &  53  &  8763.33 \\
   &  1-0  &  18  &  16244.28  &  \\
   &  1-1*  &  18  &  15275.39 \\
   &  1-2  &  21  &  14313.52 \\
   &  1-3  &  22  &  13358.69 \\
   &  1-4*  &  24  &  12410.99 \\
   &  1-5*  &  30  &  11470.63 \\
   &  1-6*  &  36  &  10537.59 \\
   &  1-7*  &  43  &  9612.56 \\
   &  2-0  &  15  &  17091.34 \\
   &  2-1  &  17  &  16122.32 \\
   &  2-2*  &  18  &  15160.33 \\
   &  2-3  &  21  &  14205.29 \\
   &  2-4  &  22  &  13257.41 \\
   &  2-5*  &  26  &  12316.62 \\
   &  2-6*  &  30  &  11383.20 \\
   &  2-7*  &  25  &  10457.20 \\
   &  3-0*  &  15  &  17932.50 \\
   &  3-1  &  14  &  16963.46 \\
   &  3-2*  &  17  &  16001.36 \\
   &  3-3*  &  17  &  15046.20 \\
   &  3-4*  &  21  &  14098.05 \\
   &  3-5  &  22  &  13157.07 \\
   &  3-6  &  27  &  12223.20 \\
   &  3-7*  &  31  &  11296.67 \\
   &  4-0*  &  14  &  18767.79 \\
   &  4-1*  &  14  &  17798.67 \\
   &  4-2  &  16  &  16836.45 \\
   &  4-3*  &  17  &  15881.27  &  \\
   &  4-4*  &  17  &  14933.00 \\
   &  4-5  &  20  &  13991.83 \\
   &  4-6  &  21  &  13057.77 \\
   &  4-7*  &  26  &  12130.78 \\
   &  5-0*  &  11  &  19597.03 \\
   &  5-1*  &  13  &  18627.83 \\
   &  5-2*  &  13  &  17665.63 \\
   &  5-3  &  14  &  16710.22 \\
   &  5-4*  &  17  &  15761.90 \\
   &  5-5*  &  20  &  14820.57 \\
   &  5-6*  &  20  &  13886.36 \\
   &  5-7  &  21  &  12959.14 \\
 \vspace{-0.5em} \\
\C{} -- \X{}  &  0-0  &  22  &  17059.99 \\
 &  0-1*  &  25  &  16091.58 \\
 &  0-2*  &  28  &  15130.40 \\
 &  0-3*  &  33  &  14176.55 \\
 &  0-4*  &  41  &  13230.35 \\
 &  0-5*  &  49  &  12292.37  \\
 &0-6*&82&11365.35\\
 &  1-0  &   &   & 17933 [1] \\
 &  2-0  &   &   &  18799 [1] \\
 &  3-0  &   &   &  19664 [1] \\
\bottomrule
\end{tabular}

[1] \cite{10BaCh.ZrO}(converted from wavelength assuming vacuum)

* Bands unobserved in rotationally resolved spectra which have been predicted by \Marvel\
\end{table}

\begin{table}
 
\centering
\caption{\label{tab:bh2}  Other singlet R-branch  bandheads in \cm{}  for \ZrO;  $J$ gives the approximate $J$ value corresponding to the rotational transitions at the bandhead.} 
\begin{tabular}{llrrrll}
\toprule
 &  v'-v"  &   $J$   &  MARVEL  &  Low-res obs. \\
\midrule
\F{} -- \X{} & 	0-0	& 17	& 25166.23 \\
& 	0-1	& 18& 	24197.07 \\
& 	0-2	& 17& 	23235.27 \\
& 	0-3	& 22& 	22280.33 \\
& 	0-4	& 22& 	21332.47 \\
& 	0-5	& 25& 	20391.58 \\
& 	0-6 & 	30	& 19458.10 \\
& 	0-7 & 	35& 	18531.99 \\
  \vspace{-0.5em} \\

\X{} -- \X{} &2-0&101&1975.73\\
&3-0&68&2917.35\\
&3-1&99&1961.43\\
&4-0&51&3858.92\\
&4-1&65&2896.33\\
&4-2&99&1948.15\\
&5-0&42&4796.37\\
&5-1&51&3831.03\\
&5-2&67&2875.26\\
&5-3&98&1932.88\\
&6-0&33&5728.41\\
&6-1&41&4761.69\\
&6-2&49&3803.14\\
&6-3&67&2854.22\\
&6-4&98&1918.42\\
&7-0&28&6654.41\\
&7-1&33&5686.80\\
&7-2&40&4726.86\\
&7-3&50&3775.20\\
  \vspace{-0.5em} \\
  \B{} -- \A{}  &  0-0  &  26  &  9507.28  &  9507.45 [1]\\
   &  0-1  &  30  &  8572.87 \\
   &  1-0  &  23  &  10359.55  &  10359.62 [1] \\
   &  1-1*  &  25  &  9424.77  &  9424.93 [1] \\
   &  2-0  &  21  &  11206.15 \\
   &  2-1  &  22  &  10271.14  &  10271.26 [1] \\
   &  3-0*  &  17  &  12047.00 \\
   &  3-1  &  21  &  11111.79 \\
   &  4-0*  &  17  &  12882.08 \\
   &  4-1*  &  17  &  11946.74 \\
   &  5-0*  &  15  &  13711.04 \\
   &  5-1*  &  17  &  12775.64 \\
\vspace{-0.5em} \\
\C{} -- \A{}  &  0-0*  &  33  &  11177.59 \\
 &  0-1*  &  41  &  10244.31 \\
 
\vspace{-0.5em} \\
 \F{} -- \A{} &0-0&22&19281.19\\
&0-1&22&18346.21\\
\vspace{-0.5em} \\

\bottomrule
\end{tabular}

[1] \cite{10BaCh.ZrO}(converted from wavelength assuming vacuum)

* Bands unobserved in rotationally resolved spectra which have been predicted by \Marvel\
\end{table}

\begin{table}
 
\centering
\caption{\label{tab:bh3}  Other singlet R-branch  bandheads in \cm{}  for \ZrO;  $J$ gives the approximate $J$ value corresponding to the rotational transitions at the bandhead.} 
\begin{tabular}{llrrlll}
\toprule
 &  v'-v"  &   $J$   &  MARVEL  \\
\midrule
\F{} -- \B{}&0-0&91&9814.07\\
&1-0&61&10637.34\\
&1-1&90&9794.47\\
\vspace{-0.5em} \\
\B{} -- \C{}&2-0&53&56.81\\
&3-0&41&893.65\\
&4-0&34&1726.14\\
&5-0&28&2553.57\\
\vspace{-0.5em} \\
\F{} -- \C{} &0-0&50&8130.59\\
&1-0&40&8961.46\\

\bottomrule
\end{tabular}
\end{table}

\begin{table}
\caption{\label{tab:bh4}  Triplet \Tb--\Ta, and \Te--\Ta R-branch  bandheads in \cm{} for \ZrO;  $J$ gives the approximate $J$ value corresponding to the rotational transitions at the bandhead.} 
\begin{tabular}{llrrlll}
\toprule
 &  v'-v"  &   $J$   &  MARVEL  &  Low-res obs. \\
\midrule
\Te$_1$ -- \X{}  &  0-0  &   &   &  19124  [1] \\
 &  1-0  &   &   &  19963 [1] \\
 &  2-0  &   &   &  20784 [1] \\
\vspace{-0.5em} \\
\Tb$_{0e}$ -- \Ta$_1$  &  0-0  &  72  &  10715.52  &  10715.26 [3] \\
 &  0-1*  &  101  &  9798.16 \\
 &  1-1  &   &   &  10634.15 [3] \\
\Tb$_{0f}$ -- \Ta$_1$  &  0-0  &  67  &  10731.97  &  10731.92 [1] \\
\vspace{-0.5em} \\
\Tb$_{1e}$ -- \Ta$_2$  &  0-0  &  63  &  10729.02  &  10728.98 [1] \\
 &  0-1*  &  89  &  9808.11 \\
\Tb$_{1f}$ -- \Ta$_2$  &  0-0  &  67  &  10731.44  &  10731.29 [1]  \\
 &  1-1  &   &   &  10649.82 [3] \\
\vspace{-0.5em} \\
\Tb$_2$ -- \Ta$_3$  &  0-0  &  62  &  10750.52  &   \\
 &  0-1  &  81  &  9828.65 \\
\midrule
\Te$_{1e}$ -- \Ta$_2$  & 0-0  &  34  &  17760.35  &  \\
 &  0-1*  &  40  &  16831.39  &  16833.0 [2] \\
 &  0-2*  &  49  &  15909.87  &  15909.2 [2] \\
 &  0-3*  &  58  &  14996.56  &  14994.3 [2] \\
\Te$_{1f}$ -- \Ta$_2$  &  0-0  &  32  &  17758.67 \\
 &  0-1*  &  38  &  16829.39  & 	16833.0 [2]\\
 &  0-2*  &  44  &  15907.41 & 	15909.2 [2]  \\
 &  0-3*  &  55  &  14993.55 	 &  14994.5 [2]\\
\Te$_{1}$ -- \Ta$_2$  &  1-1	 &   &   & 17669.2 [2] \\
 &  1-2	 &   &   & 16747.2 [2] \\
 &  1-3	 &   &   & 15556.1 [2] \\
 &  1-4	 &   &   & 14923.9 [2] \\
\bottomrule
\end{tabular}

[1] Measured \citep{81DaHa.ZrO} reassigned here, [2] \cite{88StMoKu.ZrO}, [3] \cite{10BaCh.ZrO} (converted from wavelength assuming vacuum)

* Bands unobserved in rotationally resolved spectra which have been predicted by \Marvel\
\end{table}

\begin{table*}
\centering
\caption{\label{tab:bh5}  $d ^3\Phi$--$a ^3\Delta$ R-branch  bandheads in \cm{} for \ZrO{} from the main spectroscopic networks;  $J$  gives the location of the bandhead in our data.} 
\begin{tabular}{llrrlllrrllllrrrllrr}
\toprule
 v'-v"  &   $J$   &  MARVEL  &  Low-res obs. &    $J$   &  MARVEL  &  Low-res obs. &   $J$   &  MARVEL  &  Low-res obs. \\
\midrule
 & \mc{3}{c}{\Td$_2$ -- \Ta$_1$}  &  \mc{3}{c}{\Td$_3$ -- \Ta$_2$}  &  \mc{3}{c}{\Td$_4$ -- \Ta$_3$} \\
\cmidrule(r){2-4} \cmidrule(r){5-7} \cmidrule(r){8-10}
 0-0  & 39 & 15443 &  15443.0 [1] & 34 & 15756.3 &  15756.3 [1]  & 36 & 16048.46 &  16048.6 [1] \\
 0-1  & 48 & 14515.12 &  & 43 & 14827.63 &  & 42 & 15119.3 &  15119.2 [1] \\
 0-2*   & 57 & 13595.58 &  & 52 & 13906.89 &  & 51 & 14198.01 & \\
 0-3*  & 78 & 12686.13 &  & 67 & 12995.36 &  & 68 & 13285.67 & \\
 0-4*  & 108 & 11790.97 &  & 96 & 12096.25 &  & 90 & 12385.29 & \\
 1-0  & 32 & 16290.36 &  16290.4 [1]  & 30 & 16603.14 &  16603.4 [1]  & 30 & 16897.5 &  16897.4 [1] \\
 1-1  & 37 & 15361.4 &  15361.4 [1]  & 34 & 15673.57 &  15673.5 [1] & 35 & 15967.47 &  15967.9 [1] \\
 1-2  & 44 & 14439.86 &  & 40 & 14751.22 &  & 40 & 15044.66 &  15044.0 [1] \\
 1-3*  & 57 & 12526.53 &  & 51 & 13836.77 &  & 51 & 14129.7 & \\
 1-4*  & 76 & 12623.07 &  & 68 & 12931.37 &  & 67 & 13223.55 & \\
 1-5*  & 104 & 11733.3 &  & 94 & 12047.9 &  & 87 & 12329 & \\
 2-0*  & 27 & 17132.02 &  17132.9 [1]  & 26 & 17444.03 &  17444.7 [1]  & 25 & 17741.09 & \\
 2-1  &   &   &  16202.5 [1]  & 28 & 16513.89 &  16514.0 [1]  & 29 & 16810.47 &  16810.6 [1]\\
 2-2  & 36 & 15279.76 &  15279.8 [1]  & 34 & 15590.72 &  15590.8 [1]  & 33 & 15886.85 &  15886.9 [1] \\
 2-3*  & 45 & 14364.58 &  & 40 & 14674.79 &  & 39 & 14970.55 &  14970.7 [1] \\
 2-4*  & 54 & 13457.5 &  & 51 & 13766.6 &  & 51 & 14061.86 & \\
 2-5*  & 74 & 12560.01 &  & 64 & 12867.46 &  & 64 & 13161.97 & \\
 3-0*  & 23 & 17967.59 &  & 22 & 18278.87 &  & 23 & 18579.17 & \\
 3-1*  & 26 & 17037.47 &  17036.2 [1]  & 24 & 17348.38 &  17347.7 [1]  & 25 & 17648.31 &  17646.7 [1]\\
 3-2  & 29 & 16114.24 &  16114.2 [1]  & 27 & 16424.59 &  16424.7 [1]  & 29 & 16724.12 &  16724.3 [1] \\
 3-3  & 35 & 15198.07 &  15298.4 [1]  & 34 & 15507.83 &  15508.2 [1]  & 33 & 15807.01 &  15807.6 [1] \\
 3-4  & 41 & 14289.18 &  & 39 & 14597.25 &  & 40 & 14897.02 &  14897.2 [1] \\
 3-5  & 54 & 13388.34 &  & 47 & 13696.37 &  & 51 & 13994.73 & \\
 4-0*  & 21 & 18796.88 &  & 19 & 19107.25 &  & 19 & 19411.79 & \\
 4-1*  & 24 & 17866.54 &  & 21 & 18176.52 &  & 22 & 18480.69 & \\
 4-2*  & 25 & 16942.87 &  16942.5 [1]  & 24 & 17252.48 &  17253.3 [1] & 24 & 17556.27 &  17555.6 [1] \\
 4-3  & 29 & 16026.05 &  16025.0 [1]  & 28 & 16335.14 &  16335.6 [1]  & 27 & 16638.61 &  16638.6 [1] \\ 
 4-4*  & 36 & 15116.29 &  15115.2 [1]  & 34 & 15424.86 &  15425.2 [1] & 31 & 15727.82 &  15728.3 [1] \\
 4-5  & 43 & 14213.74 &  & 39 & 14521.64 &  & 39 & 14824.26 & \\
 5-3*  &   &   &  16847.4 [1]  &  &  &  &   &   &  17463.2 [1] \\
 5-4*  &   &   &  15936.9 [1]  &  &  &  &   &   &  16554.2 [1] \\
 5-5*  &   &   &  15034.8 [1]  &   &   &  15342.6 [1]  &   &   &  15648.8 [1] \\
\bottomrule
\end{tabular}

[1] \cite{88StMoKu.ZrO} 

* Bands unobserved in rotationally resolved spectra which have been predicted by \Marvel\
\end{table*}

\subsection{Bandheads}

\Cref{tab:bh1,tab:bh2,tab:bh3,tab:bh4,tab:bh5} show bandheads
predicted by the \Marvel\ energy levels, compared with available,
low-resolution bandhead observations (note that the high resolution
bandhead observations are included in the \Marvel\ input data set).
For the singlet states, there is actually only a small number of data
points in the \B -- \A{} band that allow for direct comparison of
\Marvel{} predictions against low-resolution bandhead studies. There
are no assigned low-resolution data readily available for the \B --
\X{} band, and the low-resolution bandhead data for \C
-- \X{} does not overlap with our \Marvel{} predictions.  For the
triplets, there is good low-resolution bandhead data for the \Td --
\Ta{}, \Tb -- \Ta{} and \Te -- \Ta{} bands against which the \Marvel{}
results can be compared; in these cases, there is good agreement for
all bands, generally less than 0.5 \cm{}(though up to 1.5 \cm{}).

The \Marvel{} results provide predictions for 48 vibronic bands previously unmeasured in low or high resolution spectra. 
In contrast, there are at least further 68 low resolution bandheads whose position cannot be predicted by our \Marvel{} data due to lack of information on, usually, the excited state. The most significant missed opportunity for rotational resolved data is in the 48 non-satellite, i.e. $\Delta\Sigma=0$, \Te -- \Ta{} bandheads for $v$=0 to $v$=6 observed in low-resolution by \cite{88StMoKu.ZrO}; note that these data are not reported in this paper. Much lower resolution bandheads are identified by \cite{10BaCh.ZrO} for the \C -- \X{} band involving excited vibrational levels of the \C{} state; this too warrants further investigation to allow characterisation of the \C{} state. 

There have also been some bandheads discussed in previous \ZrO{} spectroscopic studies which our data unfortunately cannot help assign. Specifically, we don't find any recommended assignment of the double bandhead at 12082.65 \cm{}(R head) and 12069.9 \cm{}(Q head) observed by \citet{81DaHa.ZrO}. 

\subsection{Equilibrium Constants: Updated recommendations}

\begin{table*}
    \caption{Equilibrium vibrational constants, in \cm, based solely on \Marvel{} energy levels.}
    \label{tab:eqconsts}
    \centering
    \begin{tabular}{lccccccc}
    \toprule
    State  &  $T_e$  &  $\omega_e$  &  $\omega_e \chi_e$ & $B_e$ & $\alpha$(10$^{3}$)   & $D$(10$^{7}$) \\
\midrule
\X  &  0  &  976.44  &  3.45 & 0.42361 & 1.97& 3.19   \\
\A  & 5906.58  &  935.95  & - & 0.41741 & 1.89& 3.26  &  \\
\B  &  15441.70  &   859.59  &  2.99 & 0.40246 & 1.90& 3.50  \\
\F & 25229.40 & 835.24 & - & 0.39834 & 1.96 & 3.60 & \\
\midrule
\Ta$_1$  &  1099.70  &  937.74  &  3.23  & 0.41430 & 1.91 & 3.19  \\
\Ta$_2$  &  1386.90  &   938.09  &  3.24 & 0.41573 & 1.92 & 3.26   \\
 \Ta$_3$  &  1722.45  &  938.51  &  3.25 & 0.41663 & 1.93 & 3.43   \\
 [-1.5ex]\\
 \Td$_2$  &  16567.04  &  856.72  &  3.27 & 0.40419 & 2.11& 3.57  \\
 \Td$_3$  &  17169.35  &  855.84  &  3.29 & 0.40475 & 2.11& 3.61   \\
 \Td$_4$  &  17796.92  &  857.09  &  2.94 & 0.40533 & 2.07 & 3.77  \\
\bottomrule
\end{tabular}
\end{table*}

\Cref{tab:eqconsts} shows  equilibrium term energy, vibrational and rotational constants for the \X{}, \A{}, \B{}, \F{}, \Ta$_1$, \Ta$_2$, \Ta$_3$, \Td$_2$, \Td$_3$, \Td$_4$ electronic states based 
entirely on \Marvel{} energy levels. These equilibrium constants are obtained by fitting to the relevant band constants, with obvious outliers removed for averaging of $D_v$'s to obtain $D$.  Note that we have chosen to provide constants for the various sub-components of the triplet levels individually rather than use additional constants to unify their treatment. 

Based on a comprehensive collation and critical analysis of  all available information (to our knowledge) of spectroscopic constants, we can go beyond this \Marvel{} analysis to provide recommendations for all equilibrium constants for the electronic states of \ZrO{}; these are shown in \Cref{tab:UpdatedconstantsT} and \Cref{tab:UpdatedconstantsS}.  Some of these constants come solely from the \Marvel{} analysis in this paper, but some use other sources of data, particularly for vibrational anharmonicities where lower resolution bandhead data can provide additional information. Note that because we do not consider higher order corrections to $D$ or $\alpha$ within these constants, we use $D$ and $\alpha$ rather than $D_e$ and $\alpha_e$. 

\begin{table*}
\def\arraystretch{1.5}
\centering
\caption{\label{tab:UpdatedconstantsT} Recommended updated equilibrium constants in \cm{} for triplet states of \ZrO{}, with bond lengths in \AA{}. The value in the parenthesis is the uncertainty in the last figure. Justifications for each electronic state are provided in the text. }
\begin{tabular}{lrrrrrrrrrrr}
State & \mc{1}{c}{$T_e$} & \mc{1}{c}{$\omega_e$} & \mc{1}{c}{$\omega_ex_e$} & \mc{1}{c}{$B_e$} & \mc{1}{c}{$\alpha$}($10^{-3}$) &  \mc{1}{c}{$D$}($10^{-7}$) & \mc{1}{c}{$r_e$}  \\
\hline
\X{} & 0.0 & 976.44(2) & 3.44(2) & 0.4236(1) & 1.97(2) & 3.2(1) & 1.712(2) \\
\A{}  & 5906.6(2) & 942.3(2) & 3.1(1) & 0.4174(1) & 1.89(1) & 3.3(1) & 1.725(2) \\
\B{} & 15441.7(2) & 859.6(2) & 3.0(1) & 0.4025(1)& 1.90(1)  & 3.5(2) & 1.756(2) \\
\C{} & 17101(1) & 876(1) & 3.0(2) & 0.4056(1) & 1.65(1) & 3.4(1)  & 1.750(3) \\
\F{} & 25227(1) & 841(1) & 2.9(2) & 0.3983(3) & 2.0(1) & 3.6(2)  & 1.765(2) \\
\hline
\end{tabular}
\end{table*}
\begin{table*}
\def\arraystretch{1.5}
\centering
\caption{\label{tab:UpdatedconstantsS} Recommended updated equilibrium constants in \cm{} for singlet states of \ZrO{}, with bond lengths in \AA{}. Square brackets indicate the data is only from $v$=0. The value in the parenthesis is the uncertainty in the last figure. Justifications for each electronic state are provided in the text.   }
\begin{tabular}{lccccccccccc}
State & \mc{1}{c}{$T_e$ ${}_{|\Omega|=|\Lambda-1|}$}& \mc{1}{c}{$T_e$${}_{|\Omega|=|\Lambda|}$}& \mc{1}{c}{$T_e$${}_{|\Omega|=|\Lambda+1|}$} & \mc{1}{c}{$\omega_e$} & \mc{1}{c}{$\omega_ex_e$} & \mc{1}{c}{$B_e$} & \mc{1}{c}{$\alpha_e$}($10^{-3}$) & \mc{1}{c}{$D$}($10^{-7}$)  & \mc{1}{c}{$r_e$}  \\
\hline
\Ta & 1099.7(7) & 1386.9(5) & 1722.4(9)  &938.1(4) & 3.24(1) & 0.415(1) & 1.93(4) & 3.3(1) & 1.729(2)\\
\Tb & 11807(1), 11826(1)  & 12112(1) & 12469(4) & 890(1) & 3.2(3) & [0.409](1) & & 3.5(3)  & 1.741(2) \\
\Td & 16567(1) & 17169(1) & 17796(1) & 855(1) & 3.0(2) & 0.404(1) & 2.10(3)& 3.6(1)  & 1.751(2) \\
\Te &  19138(1), 19142(1) & 19177(1) & 19233(1) & 846(1) & 3.1(2) & [0.401](1) &  &  5(2) & 1.756(2) \\
\Tf &  22692(1) & 22993(1) & 23411(1)  & 821(1) & 3.3(2) & [0.392](2) &  & 3.1(6)  & 1.776(2) \\
\hline
\end{tabular}
\end{table*}

Note that these constants will provide less accurate information on particular energy levels than the raw \Marvel{} energy levels, but have the advantages of being a smaller, more easily parsed set of numbers. Thus, we have chosen to average across different parity and spin states in most cases, though we retain the term values ($T_e$) for individual spin components of the triplet states. 

The justification for each of the constants in \Cref{tab:UpdatedconstantsT} and \Cref{tab:UpdatedconstantsS} are as follows: 
\begin{itemize}
\item \underline{\X:} The  \Marvel{}  values were chosen for the main spectroscopic constants, with rounding and uncertainties determined by comparison of \Marvel{} values from \cite{76PhDaBX.ZrO} and, for rotational constants, \cite{99BeGe.ZrO}.
\item \underline{\A:} Rotational constants are from \Marvel{} analysis, while the equilibrium vibrational constants are taken from \cite{81HaDa.ZrO} (only values available). Consistency with \Marvel{} $T_v$'s has been checked. Note that \cite{81HaDa.ZrO} has rotational band constants and equilibrium vibrational constants involving \A{} $v>1$, but doesn't provide transition data involving this level thus its exclusion from the \Marvel{} compilation. 
\item \underline{\B:} Constants from \Marvel{} analysis, with uncertainties based on differences between \Marvel{} and \cite{76PhDaBX.ZrO}/\cite{81HaDa.ZrO}. Contributions from the $e$ and $f$ parity bands were averaged. 
\item \underline{\C:} Vibrational constants are taken from \cite{80Murty2.ZrO} which is based on mostly \cite{76PhDaCX.ZrO} data. $B_e$ and $\alpha_e$ were also from \cite{80Murty2.ZrO} with uncertainties chosen to ensure consistency with other available data, including \Marvel{}'s $B_0$ values. The centrifugal distortion term $D$ is by necessity a $v$=0 constant rather than an equilibrium value and thus is taken from \Marvel{} with uncertainties determined by comparison to \cite{88SiMiHu.ZrO} and \cite{76PhDaCX.ZrO}. Recommended equilibrium term energy $T_e$ is based on $T_0$ from \Marvel{} data and the adopted vibrational constants. 
\item \underline{\F:} Based on values for other states, $\omega_e x_e=2.9(2)$ seems reasonable; we use this value and other \Marvel{} $T_e$ data to obtain equilibrium term energy and vibrational constants. Rotational constants are taken from \Marvel{} values. 
\item \underline{\Ta:} \Marvel{} data is used, averaged over the various spin states for vibrational and rotational equilibrium constants. Uncertainties are estimated largely based on the difference between constants of the three different spin components. 
\item \underline{\Tb:} Rotational resolution and thus \Marvel{} data is only available for the $v$=0 levels; thus rotational $v=0$ band constants are provided rather than rotational equilibrium constants, while vibrational constants are taken from \cite{94Jonsson.ZrO}. Uncertainties in rotational band constants were estimated by comparing values from the different spin components. Uncertainties in vibrational constants were taken as 1 \cm{} based on typical differences between vibrational constants for the three spin components for \ce{ZrO} triplet states. 
\item \underline{\Td:} Constants are taken from \Marvel{} data, with uncertainties estimated based on the difference between the constants from the three different spin components. 
\item \underline{\Te:} There are no rotationally resolved $v>0$ data, so we recommend vibrational equilibrium constants from \cite{88StMoKu.ZrO} based on bandhead data. Rotational data is band constants from \Marvel{} $v$=0 levels. The equilibrium term energies, $T_e$ are calculated from the adoped equilibrium constants and \Marvel{} $T_0$ values. 
Note that there is significant enough $\Lambda$-doubling in the \Te$_0$ levels to justify separate report of different $T_e$ values, whereas this effect is negligible at the likely accuracy of these constants for the \Te$_1$ and \Te$_2$ level. 
\item \underline{\Tf:} The \cite{79HuHe} (HH) data has been retained for the vibrational equilibrium constants since there has been no subsequent experiments involving this state and no rotationally resolved spectral data for levels above $v=0$ that could be utilised in the \Marvel{} analysis. For the rotational constants, \Marvel{} data has been used, with uncertainties determined by the difference between the \Marvel{} and HH values (these are quite close) and the spread of values amongst different spin components. Note that the fitted $D_0$ constants for the \Tf$_2$ band seems erroneous and is likely the result of perturbations; thus it has been largely ignored in the averaging.  The equilibrium term energies, $T_e$, are based on \Marvel{} $T_0$ and HH vibrational constants. 
\end{itemize}

The spectroscopic constants given in \Cref{tab:UpdatedconstantsT,tab:UpdatedconstantsS} can be considered to provide a much needed update the \ZrO{} entry in the still very commonly used \cite{79HuHe} (HH) compilation of diatomic constants. 
Note that the HH data was collated up to August 1975, i.e. before a substantial number of the experiments, particularly the infrared spectra of \cite{79GaDe.ZrO} and many spectra recorded by Davis and co-authors over the 1970s and 80s. There have been significant relabelling of the electronic states over the years; we adopt the convention shown in \Cref{fig:fig1}, with some other labels, including the HH labels, shown in brackets. Our comments here use the updated notation.

A key difference between HH and our recommendations is in the harmonic vibrational frequency of the \X{} ground state: 969.7 \cm{} (HH) vs 976.38 \cm{}(\Marvel{} and our recommended value). This difference arises because the HH value is taken from a neon matrix spectrum (rather than a gas phase spectrum) which is known to cause shifts in vibrational frequencies.

All triplet states and the \A{} state harmonic
vibrational frequencies from HH were obtained from bandhead data; we
update the \A{}, \Ta{} and \Tb{} values with rotationally-resolved
data. For all states except the \X{}, \C{} and \Tb{} states, the
harmonic vibrational constants from \cite{79HuHe} are within 2-4 \cm{}
of our results. 
Our \C{} and \Tb{} vibrational constants are based on low-resolution
results from \cite{10BaCh.ZrO} and would need to be further verified;
however, they should be more reliable than those of HH.

HH does not contain any information on the observed \C{} or \Tb{} states or the theoretically predicted \D{}, \E{} and \Tc{} states. 
HH did not have access to the triplet-singlet separation, instead leaving an 'x' off-set between the singlet and triplet manifolds.  This was measured by \cite{80HaDa.ZrO} as 1100 \cm{}. The $T_e$ for the \B{}, \Ta{}, \Td{} and \Te{} states are within 2 \cm{} (\Marvel{} vs. HH). HH doesn't have absolute or relative $T_e$ for the \A{} state. 

Therefore, the key updates to HH from our results are: 
\begin{itemize}
    \item updated vibrational constants for the \X{} state;
    \item inclusion of the \Tb{}, \C{} state;
    \item absolute $T_e$ of the \A{} state;
    \item absolute $T_e$ for triplet states.
\end{itemize}

These updates are important to note given the widespread use of the HH
constants for a wide variety of applications from benchmarking quantum
chemistry \citep{90LaBa.ZrO} to calculating partition functions and
equilibrium constants for astrophysical atmosphere models
\citep{84SaTaxx.partfunc,16BaCoxx.partfunc}.

We note that we are not the first, of course, to update some of the HH constants (e.g. see \cite{87Afaf.ZrO,88DaHa.ZrO,90LaBa.ZrO}); this update is, however, comprehensive and based on a complete self-consistent data set containing all available assigned rovibronic spectra of ZrO. 


\subsection{Partition Function}


\begin{table*} 
\footnotesize
\caption{
    \label{tab:pf} Partition function for \ZrO{} as a function of
temperature($T$) estimated based on the new \Marvel{} data and reasonable extrapolations.  }
    \centering
\resizebox{\textwidth}{!}{    \begin{tabular}{lcccccccccccc}
    \toprule
 $T$ / K    &  0  &  1  &  10   &  100   &  300   &  500   &  800   &  1000   &   1500   &  2000   &  3000   &  5000  \\
\midrule
 MARVEL only          & 1. &  2.02446 &  16.8071 &  164.881 &  506.325 &  1006.32 &  2508.94 &  4185.61 &  11082.6 &  21884.9 &  53261.5 &  136797. \\
 {\bf MARVEL + constants}  &      1. &  2.02446 &  16.8071 &  164.881 &  506.398 &  1006.87 &  2510.93 &  4190.11 &  
11157.3 &  22472.4 &  59845.3 &  209393. \\

\cite{83ShLi.ZrO}  &   &   &    &   &   &   &   &  4184.00  &  11140.0  &  22450.0  &  59790.0  &  211700. \\
\cite{84SaTaxx.partfunc}  &    &   &   &    &   &    &    &  4167.99  &  11333.9  &  22729.5  &   60236.7  &  208621. \\
\cite{16BaCoxx.partfunc}  &   	1  & 	2.02843  &   	16.8283  &   	165.280  &   	507.801  &  	1010.06  &    &  	4209.23  &   	11234.5  &   	22679.4   & 	60617.8   &  	214087. \\
\bottomrule
\end{tabular}}

\vspace{2em}
\end{table*}

\Cref{tab:pf} shows the partition function for \ZrO{} at a range of
temperatures. These are predicted in two ways: using just \Marvel{} energy levels and using \Marvel{} energy levels and the contributions from rovibronic states not in the \Marvel{} collation up to $v$=15 and  $J$ =300 for the \X{}, \A{}, \B{}, \C{}, \Ta{}, \Tb{}, \Td{} and \Te{} states. We also compare against results from \cite{83ShLi.ZrO}, \cite{84SaTaxx.partfunc} and \cite{16BaCoxx.partfunc}. From these results, it is obviously essential at high temperatures to incorporate the effect of energy levels not considered in the \Marvel{} collation of energy levels (i.e. extrapolate beyond available experimental data). When this is done, the four results are all consistent within 2.6 \% at 5000 K. The key differences between the methodology for these four results are(1) explicit summation of energy levels as done in this paper vs high temperature summation expression used by previous authors,(2) the number of electronic states considered,  and(3) minor changes in the spectroscopic constants used. We have checked the convergence of the explicit sum of our partition function in terms of the values of $v$ and  $J$  and the number of electronic states included and found it to be consistent within 4 significant figures, the accuracy of our input constants, at 5000 K. Therefore, we recommend using our \Marvel{} + constants partition function values, as tabulated at 1 K intervals in the Supporting Information.



\subsection{Recommended Experiments}


It would be desirable to obtain rovibronically resolved spectra involving the higher vibrational states for the \Te, \Tb, and \C{} states (for which only $v=0$ is measured) and the \A{} and \F{} states (for which only $v=0$ and $v=1$ are measured). This is critical for a high quality spectroscopic study of the molecule; currently, line lists would need to rely on lower quality non-rotationally resolved data to understand the vibrational structure. We can use the theoretical investigation of \ZrO{} by \cite{90LaBa.ZrO} to guide our predictions for the ease of detecting these new transitions. 
The \A{} state is only reasonably accessible via relaxation or stimulated emission from the \B{} state or through high temperature initial population; several vibrational level of \B{} can be populated through observed, high intensity, transitions, however.  The \C{} state is directly accessibly from the ground \X{} state; the spectral region for the \C{} -- \X{} 1-0 transition is estimated at around 18,000 \cm{} and should have reasonable Franck-Condon intensity. Other vibronic bands of \Tb -- \Ta{} will probably be fairly weak due to near diagonal Franck-Condon factors, lower populations of vibrationally excited \Ta{} and low \Tb -- \Ta{} dipole moments. However, these bands should be detectable with few spectrally close bands interfering in absorption. 

A high resolution infrared spectrum would be desirable; the only study of \cite{79GaDe.ZrO} has very poor resolution (0.1 \cm{}).


\section{Conclusions}
We  collate all suitable available assigned \ZrO\ experimental high-resolution spectroscopy data. We  use \notrans{} assigned transitions to produce \noenergy{} energy levels in a single spectroscopic networks spanning \noelec{} electronic states and \nospinvibronic{} total spin-vibronic bands. 

The Supplementary Information to this paper contains three main files: 90Zr-16O.marvel.inp which contains the final input data of spectroscopic transitions in \Marvel\ format, 90Zr-16O.marvel.out which contains the final output energies from multiple spectroscopic networks and 90Zr-16O.energies which contains the sorted energies in the main spectroscopic network. 

Much of the data for \ZrO{} is quite outdated(for example, the \F{} state has not been investigated in more than 60 years) and would benefit from re-measurements with modern high quality techniques; it is likely some additional spin vibronic bands can be identified. However, the most pressing experimental needs for \ZrO{} are high-resolution studies of:
\begin{itemize}
\item the infrared spectra;
\item transitions that access higher vibrational levels of the \A{}, \C{} and \Tb{} state;
\item the \Te -- \X{} transitions described by \cite{10BaCh.ZrO}; this would enable another confirmation of the triplet-singlet energy separation. 
\end{itemize}
These future advances would enable significant improvements to the current understanding of the rovibronic energy-level structure of \ZrO.
New experimental data can readily be added to the existing  \Marvel{} database for \ZrO{} to produce updated empirical energy levels. These studies would substantially improve the quality of line lists for \ZrO. 

Finally we note that a major part of this work was performed by 16 and 17 year old pupils from the Highams Park
School in London, as part of a project known as ORBYTs (Original Research By Young Twinkle
Students). Three other Marvel studies were undertaken in 2016 as part of the ORBYTS project, on $^{48}$Ti$^{16}$O \citep{jt672} and the parent isotopologues of methane  \citep{jtCH4Marvel} and acetylene \citep{jt705}. Another study on H$_2$S \citep{jt718} was performed concurrently with this study in the 2016-17 academic year.  \citet{jt709} discusses our experiences of working with school students to perform high-level research.

\section*{Acknowledgments}
We would like to thank Jon Barker, Fawad Sheikh and Highams Park School for support and helpful discussions. 

We thank Bob Kurucz for providing data very quickly. 

This project has been supported by funding from the European Union 
Horizon 2020 research and innovation programme under the
Marie Sklodowska-Curie grant agreement No 701962, and by UK Science and Technology Research
Council(STFC) No. ST/M001334/1.


The work performed in Hungary was supported by the NKFIH(grant no. K119658). The collaboration between the London and Budapest teams received support from COST action CM1405, MOLIM: Molecules in Motion.

\end{document}